\documentclass[letterpaper,oneside,final,notitlepage,onecolumn]{article}
\usepackage{graphicx}
\usepackage{amsmath}
\usepackage{amsfonts}
\usepackage{amssymb}

\begin{document}

\begin{center}
{\Large HIERARCHIC THEORY OF CONDENSED MATTER \medskip}

{\Large AND ITS INTERACTION WITH LIGHT:}

{\Large \smallskip}

{\Large New Theories of Light Refraction, Brillouin Scattering\ }

{\Large and M\"{o}ssbauer effect}

\bigskip

\bigskip

\textbf{Alex Kaivarainen}\bigskip

\textbf{JBL, University of Turku, Finland, FIN-20520}

\textbf{URL: \thinspace http://www.karelia.ru/\symbol{126}alexk}

\textbf{H2o@karelia.ru\medskip}

\smallskip
\end{center}

\textbf{Materials, presented in this article are the part of new quantum
theory of condensed matter, described in:}

\textbf{[1]. Book by A. Kaivarainen: Hierarchic Concept of Matter and Field.
Water, biosystems and elementary particles. New York, 1995; ISBN 0-9642557-0-7
and two articles:}

\textbf{[2]. \thinspace''New Hierarchic Theory of Matter General for Liquids
and Solids: dynamics, thermodynamics and mesoscopic structure of water and
ice'' (see URL: http://www.karelia.ru/\symbol{126}alexk [New articles]) and:}

[\textbf{3]. \thinspace Hierarchic Concept of Matter, General for Liquids and
Solids: Water and ice (see Proceedings of the Second Annual Advanced Water
Sciences Symposium, October 4-6, 1966, Dallas, Texas.}

\textbf{Computerized verification of new described here theories are presented
on examples of WATER and ICE, using special computer program: ''Comprehensive
Analyzer of Matter Properties (CAMP)'' (copyright, 1997, A.
Kaivarainen).\bigskip}

\smallskip

\textbf{CONTENTS\smallskip}

\textbf{Summary of ''New Hierarchic Theory of Condensed Matter.''}

{\Large 1: New approach to theory of light refraction}

\textbf{1.1. Refraction in gas}

\textbf{1.2. Light refraction in liquids and solids}

{\Large 2: Mesoscopic theory of Brillouin light scattering }

\textbf{2.1. Traditional approach}

\textbf{2.2. Fine structure of scattering}

\textbf{2.3. Mesoscopic approach}

\textbf{2.4. Quantitative verification of hierarchic theory of Brillouin scattering}

{\Large 3: Mesoscopic theory of M\"{o}ssbauer effect}

\textbf{3.1. General background}

\textbf{3.2. Probability of elastic effects}

\textbf{3.3. Doppler broadening in spectra nuclear gamma-resonance (NGR)}

\textbf{3.4. Acceleration and forces, related to thermal dynamics of molecules
and ions. \thinspace Vibro-gravitational interaction}

\medskip

\begin{center}
=========================================================================

\smallskip
\end{center}

\smallskip

\begin{center}
{\large Summary of: \smallskip}

{\large \ New Hierarchic Theory of Condensed Matter}

\smallskip

{\large by: A. Kaivarainen\medskip}
\end{center}

{\large \smallskip}

\textbf{\ A basically new hierarchic quantitative theory, general for solids
and liquids, has been developed.}

\textbf{It is assumed, that unharmonic oscillations of particles in any
condensed matter lead to emergence of three-dimensional (3D) superposition of
standing de Broglie waves of molecules, electromagnetic and acoustic waves.
Consequently, any condensed matter could be considered as a gas of 3D standing
waves of corresponding nature. Our approach unifies and develops strongly the
Einstein's and Debye's models.}

\ \textbf{Collective excitations, like 3D standing de Broglie waves of
molecules, representing at certain conditions the molecular mesoscopic Bose
condensate, were analyzed, as a background of hierarchic model of condensed matter.}

\smallskip

\textbf{The most probable de Broglie wave (wave B) length is determined by the
ratio of Plank constant to the most probable impulse of molecules, or by ratio
of its most probable phase velocity to frequency. The waves B are related to
molecular translations (tr) and librations (lb).}

As the quantum dynamics of condensed matter does not follow in general case
the classical Maxwell-Boltzmann distribution, the real most probable de
Broglie wave length can exceed the classical thermal de Broglie wave length
and the distance between centers of molecules many times.

\textit{This makes possible the atomic and molecular Bose condensation in
solids and liquids at temperatures, below boiling point. It is one of the most
important results of new theory, which we have confirmed by computer
simulations on examples of water and ice.}

\smallskip

\textbf{Four strongly interrelated }new types of quasiparticles (collective
excitations) were introduced in our hierarchic model:

1.~\textit{Effectons (tr and lb)}, existing in "acoustic" (a) and "optic" (b)
states represent the coherent clusters in general case\textbf{; }

2.~\textit{Convertons}, corresponding to interconversions between \textit{tr
}and \textit{lb }types of the effectons (flickering clusters);

3.~\textit{Transitons} are the intermediate $\left[  a\rightleftharpoons
b\right]  $ transition states of the \textit{tr} and \textit{lb} effectons;

4.~\textit{Deformons} are the 3D superposition of IR electromagnetic or
acoustic waves, activated by \textit{transitons }and \textit{convertons. }

\smallskip

\ \textbf{Primary effectons }(\textit{tr and lb) }are formed by 3D
superposition of the \textbf{most probable standing de Broglie waves }of the
oscillating ions, atoms or molecules. The volume of effectons (tr and lb) may
contain from less than one, to tens and even thousands of molecules. The first
condition means validity of \textbf{classical }approximation in description of
the subsystems of the effectons. The second one points to \textbf{quantum
properties} \textbf{of coherent clusters due to molecular Bose condensation}%
\textit{. }

\ The liquids are semiclassical systems because their primary (tr) effectons
contain less than one molecule and primary (lb) effectons - more than one
molecule. \textit{The solids are quantum systems totally because both kind of
their primary effectons (tr and lb) are molecular Bose condensates.}%
\textbf{\ These consequences of our theory are confirmed by computer
calculations. }

\ The 1st order $\left[  gas\rightarrow\,liquid\right]  $ transition is
accompanied by strong decreasing of rotational (librational) degrees of
freedom due to emergence of primary (lb) effectons and $\left[
liquid\rightarrow\,solid\right]  $ transition - by decreasing of translational
degrees of freedom due to Bose-condensation of primary (tr) effectons.

\ \textbf{In the general case the effecton can be approximated by
parallelepiped with edges corresponding to de Broglie waves length in three
selected directions (1, 2, 3), related to the symmetry of the molecular
dynamics. In the case of isotropic molecular motion the effectons' shape may
be approximated by cube.}

\textbf{The edge-length of primary effectons (tr and lb) can be considered as
the ''parameter of order''.}

\smallskip

The in-phase oscillations of molecules in the effectons correspond to the
effecton's (a) - \textit{acoustic }state and the counterphase oscillations
correspond to their (b) - \textit{optic }state. States (a) and (b) of the
effectons differ in potential energy only, however, their kinetic energies,
impulses and spatial dimensions - are the same. The \textit{b}-state of the
effectons has a common feature with \textbf{Fr\"{o}lich's polar mode. }

\smallskip

\textbf{The }$(a\rightarrow b)$\textbf{\ or }$(b\rightarrow a)$%
\textbf{\ transition states of the primary effectons (tr and lb), defined
as\ primary transitons, are accompanied by a change in molecule polarizability
and dipole moment without density fluctuations. At this case they lead to
absorption or radiation of IR photons, respectively.}

\textbf{\ Superposition (interception) of three internal standing IR photons
of different directions (1,2,3) - forms primary electromagnetic deformons (tr
and lb).}

\ On the other hand, the [lb$\rightleftharpoons\,$tr] \textit{convertons }and
\textit{secondary transitons} are accompanied by the density fluctuations,
leading to \textit{absorption or radiation of phonons}.

\textit{Superposition resulting from interception} of standing phonons in
three directions (1,2,3), forms \textbf{secondary acoustic deformons (tr and
lb). }

\smallskip

\ \textit{Correlated collective excitations }of primary and secondary
effectons and deformons (tr and lb)\textbf{, }localized in the volume of
primary \textit{tr }and \textit{lb electromagnetic }deformons\textbf{, }lead
to origination of \textbf{macroeffectons, macrotransitons}\textit{\ }and
\textbf{macrodeformons }(tr and lb respectively)\textbf{. }

\ \textit{Correlated simultaneous excitations of \thinspace tr and lb}
\textit{macroeffectons }in the volume of superimposed \textit{tr }and
\textit{lb }electromagnetic deformons lead to origination of
\textbf{supereffectons. }

\ In turn, the coherent excitation of \textit{both: tr }and \textit{lb
macrodeformons and macroconvertons }in the same volume means creation of
\textbf{superdeformons.} Superdeformons are the biggest (cavitational)
fluctuations, leading to microbubbles in liquids and to local defects in solids.

\smallskip

\ \textbf{Total number of quasiparticles of condensed matter equal to 4!=24,
reflects all of possible combinations of the four basic ones [1-4], introduced
above. This set of collective excitations in the form of ''gas'' of 3D
standing waves of three types: de Broglie, acoustic and electromagnetic - is
shown to be able to explain virtually all the properties of all condensed matter.}

\ \textit{The important positive feature of our hierarchic model of matter is
that it does not need the semi-empiric intermolecular potentials for
calculations, which are unavoidable in existing theories of many body systems.
The potential energy of intermolecular interaction is involved indirectly in
dimensions and stability of quasiparticles, introduced in our model.}

{\large \ The main formulae of theory are the same for liquids and solids and
include following experimental parameters, which take into account their
different properties:}

$\left[  1\right]  $\textbf{- Positions of (tr) and (lb) bands in oscillatory spectra;}

$\left[  2\right]  $\textbf{- Sound velocity; }$\,$

$\left[  3\right]  $\textbf{- Density; }

$\left[  4\right]  $\textbf{- Refraction index (extrapolated to the infinitive
wave length of photon}$)$\textbf{.}

\textit{\ The knowledge of these four basic parameters at the same temperature
and pressure makes it possible using our computer program, to evaluate more
than 300 important characteristics of any condensed matter. Among them are
such as: total internal energy, kinetic and potential energies, heat-capacity
and thermal conductivity, surface tension, vapor pressure, viscosity,
coefficient of self-diffusion, osmotic pressure, solvent activity, etc. Most
of calculated parameters are hidden, i.e. inaccessible to direct experimental measurement.}

\ The new interpretation and evaluation of Brillouin light scattering and
M\"{o}ssbauer effect parameters may also be done on the basis of hierarchic
theory. Mesoscopic scenarios of turbulence, superconductivity and superfluity
are elaborated.

\ Some original aspects of water in organization and large-scale dynamics of
biosystems - such as proteins, DNA, microtubules, membranes and regulative
role of water in cytoplasm, cancer development, quantum neurodynamics, etc.
have been analyzed in the framework of Hierarchic theory.

\medskip

\textbf{Computerized verification of our Hierarchic concept of matter on
examples of water and ice is performed, using special computer program:
Comprehensive Analyzer of Matter Properties (CAMP, copyright, 1997,
Kaivarainen). The new opto-acoustical device (CAMP), based on this program,
with possibilities much wider, than that of IR, Raman and Brillouin
spectrometers, has been proposed}

\textbf{\ (see URL:\thinspace\thinspace http://www.karelia.ru/\symbol{126}alexk).}

\smallskip

\textbf{This is the first theory able to predict all known experimental
temperature anomalies for water and ice. The conformity between theory and
experiment is very good even without any adjustable parameters. }

\textbf{The hierarchic concept creates a bridge between micro- and macro-
phenomena, dynamics and thermodynamics, liquids and solids in terms of quantum
physics. }\newpage

\smallskip

\begin{center}
{\Large \smallskip1: New approach to theory of light refraction}

{\large 1.1. Refraction in gas}
\end{center}

\smallskip

\textbf{If the action of photons onto electrons of molecules is considered as
a force, activating a harmonic oscillator with decay, it leads to the known
classical equations for a complex refraction index (Vuks, 1984).}

\textbf{The Lorentz-Lorenz formula obtained in such a way is convenient for
practical needs. However, it does not describe the dependence of refraction
index on the incident light frequency and did not take into account the
intermolecular interactions. In the new theory proposed below we have tried to
clear up the relationship between these parameters.}

\textbf{Our basic idea is that the dielectric penetrability of matter
}$\epsilon$\textbf{, (equal in the optical interval of frequencies to the
refraction index squared }$\;n^{2})$\textbf{, is determined by the ratio of
partial volume energies of photon in vacuum to similar volume energy of photon
in matter:}%

\begin{equation}
\epsilon=n^{2}={\frac{[E_{p}^{0}]}{[E_{p}^{m}]}}={\frac{m_{p}c^{2}}{m_{p}%
c_{m}^{2}}}={\frac{c^{2}}{c_{m}^{2}}}\tag{1.1}%
\end{equation}

where $m_{p}=h\nu_{p}/c^{2}$ is the effective photon mass, $\,\,c\,$ is the
light velocity in vacuum, $c_{m}$ is the effective light velocity in matter.

\textbf{We introduce the notion of partial volume energy of a photon in vacuum
}$[E_{p}^{0}]$\textbf{\ and in matter }$[E_{p}^{m}]$\textbf{\ as a product of
photon energy }$(E_{p}=h\nu_{p})$\textbf{\ and the volume }$(V_{p}%
)$\textbf{\ occupied by 3D standing wave of photon in vacuum and in matter, correspondingly:}%

\begin{equation}
\lbrack E_{p}^{0}]=E_{p}V_{p}^{0}\qquad[E_{p}^{m}]=E_{p}V_{p}^{m}\tag{1.2}%
\end{equation}

The 3D standing photon volume as an interception volume of 3 different
standing photons normal to each other was termed in our mesoscopic model as a
primary electromagnetic deformon (see Introduction of [1,2,3]).

In vacuum, where the effect of an \textbf{excluded volume due to the spatial
incompatibility of electron shells of molecules and photon }is absent, the
volume of $\,3D$ photon standing wave (primary deformon) is:%

\begin{equation}
V_{p}^{0}={\frac{1}{n_{p}}}={\frac{3\lambda_{p}^{2}}{8\pi}}\tag{1.3}%
\end{equation}
\textbf{We will consider the interaction of light with matter in this
mesoscopic volume, containing a thousands of molecules of condensed matter. It
is the reason why we titled this theory of light refraction as mesoscopic one.}

Putting (1.3) into (1.2), we obtain the formula for the partial volume energy
of a photon in vacuum:%

\begin{equation}
\lbrack E_{p}^{0}]=E_{p}V_{p}^{0}=h\nu_{p}{\frac{9\lambda_{p}^{2}}{8\pi}%
}={\frac{9}{4}}\hbar c\lambda_{p}^{2}\tag{1.4}%
\end{equation}

Then we proceed from the assumption that waves B of photons can not exist with
waves B of electrons, forming the shells of atoms and molecules in the same
space elements. Hence, the effect of excluded volume appears during the
propagation of an external electromagnetic wave through the matter. It leads
to the fact that in matter the volume occupied by a photon, is equal to%

\begin{equation}
V_{p}^{m}=V_{p}^{0}-V_{p}^{\text{ex}}=V_{p}^{0}-n_{M}^{p}\cdot V_{e}%
^{M}\tag{1.5}%
\end{equation}
where $V_{p}^{\text{ex}}=n_{M}^{p}V_{e}^{M}$ is the excluded volume which is
equal to the product of the number of molecules in the volume of one photon
standing wave $(n_{M}^{p})$ and the volume occupied by the electron shell of
one molecule $(V_{e}^{M})$.

$n_{M}^{p}$ is determined by the product of the volume of the photons 3D
standing wave in the vacuum (1.3) and the concentration of molecules
$(n_{M}=N_{0}/V_{0})$:%

\begin{equation}
n_{M}^{p}={\frac{9\lambda_{p}^{3}}{8\pi}}\left(  {\frac{N_{0}}{V_{0}}}\right)
\tag{1.6}%
\end{equation}
In the absence of the polarization by the external field and intermolecular
interaction, the volume occupied by electrons of the molecule:%

\begin{equation}
V_{e}^{M}={\frac{4}{3}}\pi L_{e}^{3}\tag{1.7}%
\end{equation}
where $L_{e}$ is the radius of the most probable wave $B\;(L_{e}=\lambda
_{e}/2\pi)$ of the outer electron of a molecule. As it has been shown in (7.5)
that the mean molecule polarizability is:%

\begin{equation}
\alpha=L_{e}^{3}\tag{1.8}%
\end{equation}
Then taking (1.7) and (1.6) into account, the excluded volume of primary
electromagnetic deformon in the matter is:%

\begin{equation}
V_{p}^{\text{ex}}={\frac{9\lambda_{p}^{3}}{8\pi}}n_{M}{\frac{4}{3}}\pi
\alpha={\frac{3}{2}}\lambda_{p}^{3}n_{M}\alpha\tag{1.9}%
\end{equation}
Therefore, the partial volume energy of a photon in the vacuum is determined
by eq.(1.4), while that in matter, according to (1.5):%

\begin{equation}
\lbrack E_{p}^{m}]=E_{p}\cdot V_{p}^{m}=E_{p}\cdot[V_{p}^{0}-V_{p}^{\text{ex}%
}]\tag{1.10}%
\end{equation}
Putting (1.4) and (1.10) into (1.1) we obtain:%

\begin{equation}
\epsilon=n^{2}={\frac{E_{p}V_{p}^{0}}{E_{p}(V_{p}^{0}-V_{p}^{\text{ex}})}%
}\tag{1.11}%
\end{equation}
or

\begin{quotation}%
\begin{equation}
{\frac{1}{n^{2}}}=1-{\frac{V_{p}^{\text{ex}}}{V^{0}}}\tag{1.12}%
\end{equation}
\textbf{Then, putting eq.(1.9) and (1.3) into (1.12) we derive new equation
for refraction index, leading from our mesoscopic theory:}%

\begin{equation}
{\frac{1}{n^{2}}}=1-{\frac{4}{3}}\pi n_{M}\alpha\tag{1.13}%
\end{equation}
\textbf{or in another form:}
\end{quotation}%

\begin{equation}
{\frac{n^{2}-1}{n^{2}}}={\frac{4}{3}}\pi n_{M}\alpha={\frac{4}{3}}\pi
{\frac{N_{0}}{V_{0}}}\alpha\tag{1.14}%
\end{equation}

\textbf{where: }$n_{M}=N_{0}/V_{0}$\textbf{\ is a concentration of molecules;}

\textbf{\smallskip}

\textbf{In this equation }$\alpha=L_{e}^{3}$\textbf{\ is the average static
polarizability of molecules for the case when the external electromagnetic
fields as well as intermolecular interactions inducing the additional
polarization are absent. This situation is realized at }$\,E_{p}=h\nu
_{p}\rightarrow0$\textbf{\ \thinspace and }$\,\lambda_{p}\rightarrow\infty\,
$\textbf{\ in the gas phase. As will be shown below the value of resulting
}$\alpha^{*}\,\,$\textbf{in condensed matter is bigger.}

\textbf{\smallskip}

\begin{center}
{\large 1.2. Light refraction in liquids and solids}
\end{center}

\smallskip

According to the Lorentz classical theory, the electric component of the outer
electromagnetic field is amplified by the additional inner field
$(E_{\text{ad}})$, related to the interaction of induced dipole moments in
composition of condensed matter with each other:
\begin{equation}
E_{\text{ad}}={\frac{n^{2}-1}{3}}E\tag{1.15}%
\end{equation}
The mean Lorentz \textit{acting field }$\bar{F}$ can be expressed as:%

\begin{equation}
\overline{F}=E+E_{\text{ad}}={\frac{n^{2}+2}{3}}E\;\;\;\;(\text{at
\ }n\rightarrow1,\;\,\overline{F}\rightarrow E)\tag{1.16}%
\end{equation}
$\;\;\;\;\,\,\bar{F}$- has a dimensions of electric field tension and tends to
E in the gas phase when $n\rightarrow1$.

\smallskip

\textbf{In accordance with our mesoscopic model, except the Lorentz acting
field, the total internal acting field, includes also two another
contributions, increasing the molecules polarizability (}$\alpha$\textbf{) in
condensed matter:}

\textbf{1.~Potential intermolecular field, including all the types of Van-
der-Waals interactions in composition of coherent collective excitations, even
without external electromagnetic field. Like total potential energy of matter,
this contribution must be dependent on temperature and pressure;}

\textbf{2.~Primary internal field, related with primary electromagnetic
deformons (tr and lib). This component of the total acting field also exist
without external fields. Its frequencies corresponds to IR range and its
action is much weaker than the action of the external visible light.}

\smallskip

Let us try to estimate the energy of the total acting field and its effective
frequency ($\nu_{f}$) and wavelength $(\lambda_{f})$, that we introduce as:%

\begin{equation}
A_{f}=h\nu_{f}={\frac{hc}{\lambda_{f}}}=A_{L}+A_{V}+A_{D}\tag{1.17}%
\end{equation}

where: $A_{L},\,\,\,A_{V}$ and $A_{D}$ are contributions, related with Lorentz
field, potential field and primary deformons field correspondingly.

\smallskip

When the interaction energy of the molecule with a photon $(E_{p}=h\nu_{p})$
is less than the energy of the resonance absorption, then it leads to elastic
polarization of the electron shell and origination of secondary photons, i.e.
light scattering. We assume in our consideration that the increment of
polarization of a molecule $(\alpha)$ under the action of the external photon
$(h\nu_{p})$ and the total active field $(A_{f}=h\nu_{f})$ can be expressed
through the increase of the most probable radius of the electron's shell
$(L_{e}=\alpha^{1/3})$, using our (eq. 7.6 from [1]):
\begin{equation}
\Delta L_{e}={\frac{\omega_{p}\,\,m_{e}}{2\hbar}}\alpha\tag{1.18}%
\end{equation}

where the resulting increment:%

\begin{equation}
\Delta L^{*}=\Delta L_{e}+\Delta L_{f}={\frac{(h\nu_{p}+A_{f})m_{e}}%
{2\hbar^{2}}}\alpha\tag{1.18a}%
\end{equation}

where: $\alpha=L_{e}^{3}$ is the average polarizability of molecule in gas
phase at $\nu_{f}\rightarrow$ 0.

For water molecule in the gas phase:
\[
L_{e}=\alpha^{1/3}=1.13\cdot10^{-10}\,m
\]
\ \ is a known constant, determined experimentally [4].

\textbf{The total increment of polarizability radius }$(\Delta L^{*}%
)$\textbf{\ and resulting polarizability of molecules (}$\alpha^{*}%
)\,\,\,$\textbf{in composition of condensed matter affected by the acting
field}
\[
\mathbf{\ }\alpha^{*}=(L^{*})^{3}
\]
\textbf{can be find from the experimental refraction index (n) using our
formula (1.14):}%

\begin{equation}
L^{*}=\left[  {\frac{3}{4\pi}}{\frac{V_{0}}{N_{0}}\cdot}{\frac{n^{2}-1}{n^{2}%
}}\right]  ^{1/3}\tag{1.19}%
\end{equation}%
\begin{equation}
\Delta L^{*}=L^{*}-L_{e}\tag{1.20}%
\end{equation}%
\[
\text{where: \ }L^{*}=(\alpha^{*})^{1/3}
\]
From (1.18) we get a formula for the increment of radius of polarizability
$(\Delta L_{f})$, induced by the total internal acting field:%

\begin{equation}
\Delta L_{f}=\Delta L^{*}-\Delta L_{e}={\frac{A_{f}\cdot m_{e}}{2\hbar^{2}}%
}\alpha\tag{1.21}%
\end{equation}

Like total internal acting field energy (1.17), this total acting increment
can be presented as a sum of contributions, related to Lorentz field
$\;(\Delta L_{F})$, \thinspace potential field $(\Delta L_{V})$ and primary
deformons field $(\Delta L_{D})$:%

\begin{equation}
\Delta L_{f}=\Delta L_{L}+\Delta L_{V}+\Delta L_{D}\tag{1.22}%
\end{equation}
Increment $\Delta L_{e}$, induced by external photon only, can be calculated
from the known frequency ($\nu_{p}$) of the incident light (see 1.18a):%

\begin{equation}
\Delta L_{e}={\frac{h\nu_{p}\cdot m_{e}}{2\hbar^{2}}}\alpha\tag{1.23}%
\end{equation}

It means that $\Delta L_{f}$ \thinspace can be found from (1.21) and (1.17),
using (1.23). Then from (1.21) we can calculate the energy $(A_{f})$,
effective frequency $(\nu_{f})$ and wave length $(\lambda_{f})$ of the total
acting field like:%

\begin{equation}
A_{f}=h\nu_{f}=hc/\lambda_{f}=2\frac{\Delta L_{f}\cdot\hbar^{2}}{m_{e}\alpha
}\tag{1.24}%
\end{equation}
The computer calculations of $\alpha^{*};\;L^{*}=L_{e}+\Delta L^{*}%
=(\alpha^{*})^{1/3}\;\,$and$\,\;\;A_{f}$ \thinspace\ in the temperature range
$(0-95^{0})$ are presented on Fig.1.1.

One must keep in mind that in general case $\alpha$ and $L$ are tensors. It
means that all the increments, calculated on the base of eq.(1.18a) must be
considered as the effective ones. Nevertheless, it is obvious that our
approach to analysis of the acting field parameters can give useful additional
information about the properties of transparent condensed matter.

\begin{center}%
\begin{center}
\includegraphics[
height=2.3557in,
width=4.8879in
]%
{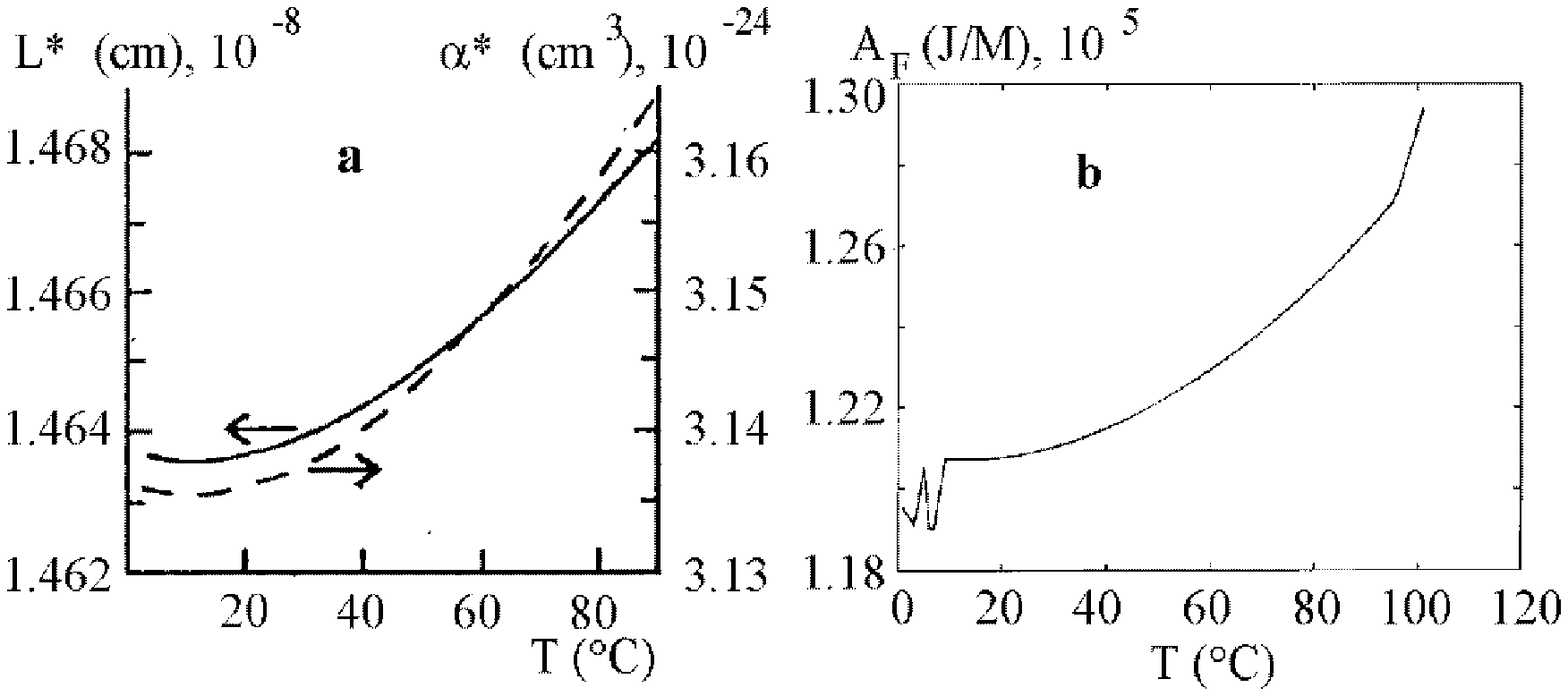}%
\end{center}
\end{center}

\begin{quotation}
\smallskip

\textbf{Fig.~1.1. }(a)- Temperature dependencies of the most probable outer
electron shell radius of $H_{2}O\;(L^{*})$ and the effective polarizability
$\alpha^{*}=(L^{*})^{3}$ in the total acting field;

(b)- Temperature dependence of the total acting field $(A_{f})$ energy in
water at the wavelength of the incident light $\lambda_{p}=5.461\cdot
10^{-5}cm^{-1}$. The experimental data for refraction index n(t) were used in
calculations. The initial electron shell radius is: $L_{e}=\alpha_{H2O}%
^{1/3}=1.13\cdot10^{-8}cm\;$[4]. \ In graphical calculations in Fig.1.1a, the
used experimental temperature dependence of the water refraction index were
obtained by Frontas'ev and Schreiber [5].
\end{quotation}

\medskip

\textbf{The temperature dependencies of these parameters were computed using
the known experimental data on refraction index }$n(t)$\textbf{\ for water and
presented in Fig.1.1a. The radius }$L^{\ast}$\textbf{\ in the range }%
$0-95^{0}C$\textbf{\ increases less than by 1\% at constant incident light
wavelength }$(\lambda=546.1nm)$\textbf{. The change of }$\Delta L_{f}%
$\textbf{\ with temperature is determined by its potential field component
change}$\;\Delta L_{V}$\textbf{.}

\textbf{The relative change of this component: }$\Delta\Delta L_{V}/\Delta
L_{f}\;(t=0^{0}C)$\textbf{\ is about 9\%. Corresponding to this change the
increasing of the acting field energy }$A_{f}\;$\textbf{(eq.1.23) increases
approximately by }$8\,kJ/M\;($\textbf{Fig 1.1 b) due to its potential field contribution.}

\textbf{It is important that the total potential energy of water in the same
temperature range, according to our calculations, increase by the same
magnitude (Fig.5b in [1] or Fig.3b in [ 3]). This fact points to the strong
correlation between potential intermolecular interaction in matter and the
value of the acting field energy.}

\textbf{It was calculated that, at constant temperature }$(20^{0}%
)$\textbf{\ the energy of the acting field }$(A_{f}),\,(eq.1.23)$\textbf{\ in
water practically does not depend on the wavelength of incident light
}$(\lambda_{p})$\textbf{. At more than three time alterations of }$\lambda
_{p}$\textbf{: from }$12.56\cdot10^{-5}cm\,$\textbf{\ to\thinspace}%
$3.03\cdot10^{-5}cm$\textbf{\ and the water refraction index }$\left(
n\right)  $\textbf{\ from 1.320999 to 1.358100 [6] the value of }$A_{f}%
$\textbf{\ changes less than by 1\%.}

\textbf{At the same conditions the electron shell radius L}$^{*}$\textbf{\ and
the acting polarizability }$\alpha^{*}$\textbf{\ thereby increase from (1.45
to 1.5)}$\cdot$\textbf{10}$^{-10}\,m$\textbf{\ and from (3.05 to }%
$3.274)\cdot10^{-30}m^{3}$\textbf{\ respectively (Fig.1.2). These changes are
due to the incident photons action only. For water molecules in the gas phase
and }$\lambda_{p}\rightarrow\infty$\textbf{\ the initial polarizability
}$(\alpha=L_{e}^{3})$\textbf{\ is equal to }$1.44\cdot10^{-24}cm^{3}%
\;$\textbf{[4], i.e. significantly less than in condensed matter under the
action of external and internal fields.}

\textbf{Obviously, the temperature change of energy }$A_{f}$%
\textbf{\ (Fig.1.1b) is determined by the internal pressure increasing
(section 11.2 of [1]), related to intermolecular interaction change, depending
on mean distances between molecules and, hence, on the concentration }%
$(N_{0}/V_{0})$\textbf{\ of molecules in condensed matter.}

\begin{center}
\medskip

\smallskip%
\begin{center}
\includegraphics[
height=2.5538in,
width=5.2918in
]%
{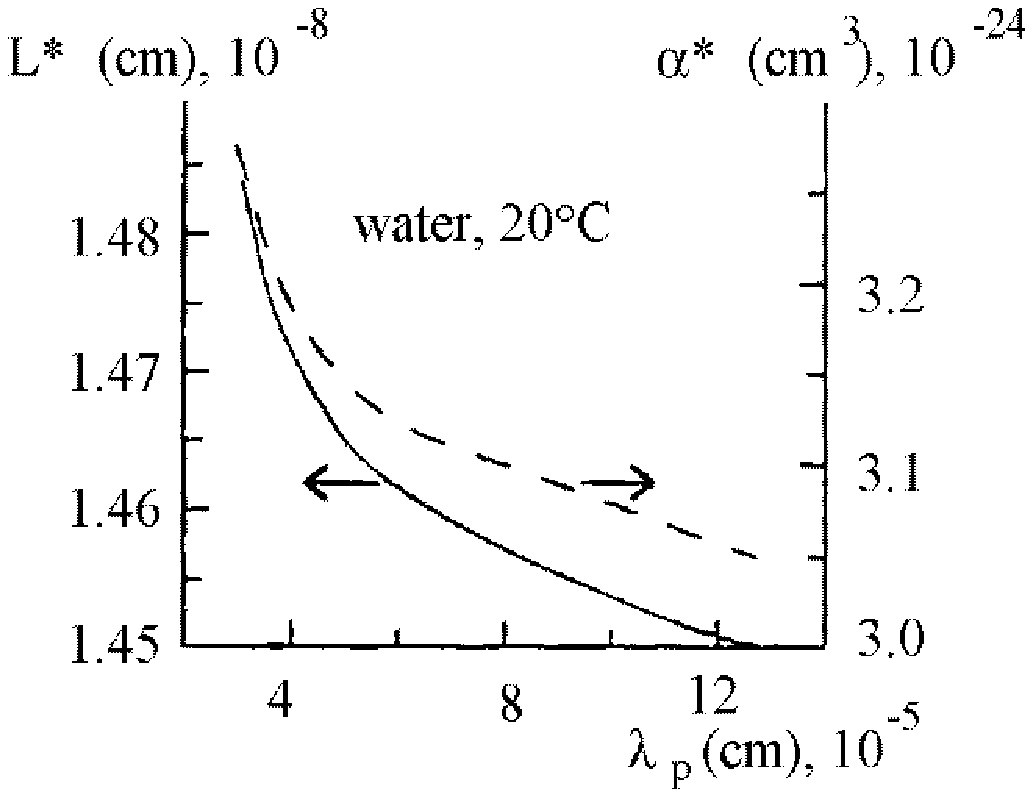}%
\end{center}
\end{center}

\begin{quotation}
\textbf{Fig.~1.2. }Dependencies of the acting polarizability $\alpha
^{*}=(L^{*})^{3}$ and electron shell radius of water in the acting field
$(L^{*})$ on incident light wavelength $(\lambda_{p})$, calculated from eq.
(1.14) and experimental data $n(\lambda_{p})\;[6]$.\thinspace\thinspace The
initial polarizability of $\,H_{2}O$ in the gas phase at $\,\lambda
_{p}\rightarrow\infty$ is equal to $\,\alpha=L_{e}^{3}=1.44\cdot10^{-24}%
cm^{3}$. The corresponding initial radius of the $H_{2}O$ electron shell
is$\;L_{e}=1.13\cdot10^{-8}cm$.

\medskip
\end{quotation}

\textbf{On the basis of our data, changes of }$A_{f},$\textbf{\ calculated
from (1.24) are caused mainly by the heat expansion of the matter. The photon
induced increment of the polarizability }$(\alpha\rightarrow\alpha^{*}%
)$\textbf{\ practically do not change }$A_{f}.$

\textbf{The ability to obtain new valuable information about changes of
molecule polarizability under the action of incident light and about
temperature dependent molecular interaction in condensed medium markedly
reinforce such a widely used method as refractometry.}

\textbf{The above defined relationship between the molecule polarizability and
the wave length of the incident light allows to make a new endeavor to solve
the light scattering problems.}

\bigskip

\begin{center}
\bigskip

{\Large 2.\ Mesoscopic theory of Brillouin light scattering in condensed matter}

\smallskip

{\large 2.1. Traditional approach}
\end{center}

\smallskip

According to traditional concept, light scattering in liquids and crystals as
well as in gases takes place due to random heat fluctuations. In condensed
media the fluctuations of density, temperature and molecule orientation are possible.

Density ($\rho$) fluctuations leading to dielectric penetrability $(\epsilon)$
fluctuations are of major importance. This contribution is estimated by means
of Einstein formula for scattering coefficient of liquids [7]:%

\begin{equation}
R={\frac{Ir^{2}}{I_{0}V}}={\frac{\pi}{2\lambda^{4}}}kT\beta_{T}\left(
\rho{\frac{\partial\epsilon}{\partial\rho}}\right)  _{T}\tag{2.1}%
\end{equation}

where $\beta_{T}$ is isothermal compressibility.

Many authors made attempts to find a correct expression for the variable
$(\rho{\frac{\partial\epsilon}{\partial\rho}).}$

The formula derived by Vuks [8, 9] is most consistent with experimental data:%

\begin{equation}
\rho{\frac{\partial\epsilon}{\partial\rho}}=(n^{2}-1){\frac{3n^{2}}{2n^{2}-1}%
}\tag{2.2}%
\end{equation}

\smallskip

\begin{center}
{\large 2.2. Fine structure of scattering}
\end{center}

\smallskip

The fine structure - spectrum of the scattering in liquids is represented by
two Brillouin components with frequencies shifted relatively from the incident
light frequency: $\nu_{\pm}=\nu_{0}\pm\Delta\nu$ and one unshifted band like
in gases $(\nu_{0})$.

The shift of the Brillouin components is caused by the Doppler effect
resulting from a fraction of photons scattering on phonons moving at sound
speed in two opposite directions [8].

This shift can be explained in different way as well. If in the antinodes of
the standing wave the density oscillation occurs at frequency ($\Omega$):%

\begin{equation}
\rho=\rho_{0}\cos\Omega t,\tag{2.3}%
\end{equation}
then the scattered wave amplitude will change at the same frequency. Such a
wave can be represented as a superposition of two monochromatic waves having
the frequencies:$(\omega+\Omega)$ and $(\omega-\Omega)$, where%

\begin{equation}
\Omega=2\pi f\tag{2.4}%
\end{equation}
is the elastic wave frequency at which scattering occurs when the Wolf-Bragg
condition is satisfied:%

\begin{equation}
2\Lambda\sin\varphi=2\Lambda\sin{\frac{\theta}{2}}=\lambda^{^{\prime}{}%
}\tag{2.5}%
\end{equation}
or%

\begin{equation}
\Lambda=\lambda^{\prime}{}/(2\sin{\frac{\theta}{2}})={\frac{c}{n\nu}}%
(2\sin{\frac{\theta}{2}})=v_{ph}/f\tag{2.6}%
\end{equation}
where $\Lambda$ is the elastic wave length corresponding to the frequency
$f;\;\lambda^{^{\prime}{}}=\lambda/n=c/n\nu\;(\lambda^{\prime}{}$ and
$\lambda$ are the incident light wavelength in matter and vacuum,
respectively$);\;\varphi$ is the angle of sliding; $\theta$ is the angle of
scattering; n is the refraction index of matter; $c\,\,\,$is the light speed.

The value of Brillouin splitting is represented as:%

\begin{equation}
\pm\Delta\nu_{M-B}=f={\frac{V_{ph}}{\Lambda}}=2\nu{\frac{V_{ph}}{c}}%
n\sin{\frac{\theta}{2}}\tag{2.7}%
\end{equation}

where:\ $\nu n/c=1/\lambda;\;n$ is the refraction index of matter; $\;\nu$ is
incident light frequency;%

\begin{equation}
v_{ph}=v_{S}\tag{2.8}%
\end{equation}
is the phase velocity of a scattering wave equal to hypersonic velocity.

The formula (2.7) is identical to that obtained from the analysis of the
Doppler effect:%

\begin{equation}
{\frac{\Delta\nu}{\nu}}=\pm2{\frac{V_{S}}{c}}n\sin{\frac{\theta}{2}}\tag{2.9}%
\end{equation}

According to the classical theory, the central line, which is analogous to
that observed in gases, is caused by entropy fluctuations in liquids, without
any changes of pressure [8]. On the basis of Frenkel theory of liquid state,
the central line can be explained by fluctuations of ''hole'' number -
cavitational fluctuations [10].

The thermodynamic approach of Landau and Plachek leads to the formula, which
relates the intensities of the central (I) and two lateral $(I_{M-B})$ lines
of the scattering spectrum with compressibility and heat capacities:%

\begin{equation}
{\frac{I}{2I_{M-B}}}={\frac{I_{p}}{I_{\text{ad}}}}={\frac{\beta_{T}-\beta_{S}%
}{\beta_{S}}}={\frac{C_{p}-C_{v}}{C_{v}}}\tag{2.10}%
\end{equation}
where: $\beta_{T}\,$ and\thinspace$\beta_{S}$ are isothermal and adiabatic
compressibilities; $C_{p}$ and $\,C_{v}$ are isobaric and isohoric heat capacities.

In crystals, quartz for example, the central line in the fine structure of
light scattering is usually absent or very small. However, instead of one pair
of shifted components, observed in liquids, there appear \textit{three
}Brillouin components in crystals. One of them used to be explained by
scattering on the longitudinal phonons, and two - by scattering on the
transversal phonons.

\smallskip

\begin{center}
{\large 2.3. New mesoscopic approach to problem}
\end{center}

\smallskip

\textbf{In our hierarchic theory the thermal ''random'' fluctuations are
''organized'' by different types of superimposed quantum excitations.}

\textbf{According to our Hierarchic model, including microscopic, mesoscopic
and macroscopic scales of matter (see Introduction of [1,2,3]), the most
probable (primary) and mean (secondary) effectons, translational and
librational are capable of quantum transitions between two discreet states:
}$(a\Leftrightarrow b)_{tr,lb}$\textbf{\ and }$(\bar{a}\Leftrightarrow\bar
{b})_{tr,lb}$\textbf{\ respectively. These transitions lead to
origination/annihilation of photons and phonons, forming primary and secondary deformons.}

\textbf{The mean heat energy of molecules is determined by the value of 3kT,
which as our calculations show, has the intermediate value between the
discreet energies of a and b quantum states of primary effectons (Fig.19 of
[1]), making, consequently, the non equilibrium conditions in condensed
matter. Such kind of instability is a result of ''competition'' between
classical and quantum distributions of energy .}

The maximum deviations from thermal equilibrium and that of the dielectric
properties of matter occur when the same states of primary and secondary
quasiparticles, e.g. \textit{a},\textit{\={a} }and \textit{b},\textit{\={b}
}occur simultaneously. Such a situation corresponds to the A and B states of
macroeffectons. The $(A\Leftrightarrow B)_{tr,lb}$ transitions represent
thermal fluctuations. The big density fluctuations are related to ''flickering
clusters'' (macroconvertions between librational and translational primary
effectons) and the maximum fluctuations correspond to Superdeformons.

\textbf{Only in the case of spatially independent fluctuations the
interference of secondary scattered photons does not lead to their total compensation.}

The probability of the event that two spatially uncorrelated events coincide
in time is equal to the product of their independent probabilities [10].

Thus, the probabilities of the coherent (\textit{a},\textit{\={a}}) and
(\textit{b},\textit{\={b}}) states of primary and secondary effectons,
corresponding to A and B states of the macroeffectons (tr and lib),
independent on each other, are equal to:%

\begin{equation}
\left(
\begin{array}
[c]{c}%
P_{M}^{A}%
\end{array}
\right)  _{tr,lb}^{\text{ind}}=\left(
\begin{array}
[c]{c}%
P_{ef}^{a}\bar{P}_{ef}^{a}%
\end{array}
\right)  _{tr,lb}^{S}\cdot\left(  {\frac{1}{Z^{2}}}\right)  =\left(
{\frac{P_{M}^{A}}{Z^{2}}}\right)  _{tr,lb}\tag{2.11}%
\end{equation}%
\begin{equation}
\left(  P_{M}^{B}\right)  _{tr,lb}^{\text{ind}}=\left(
\begin{array}
[c]{c}%
P_{ef}^{b}\bar{P}_{ef}^{b}%
\end{array}
\right)  _{tr,lb}^{S}\cdot\left(  {\frac{1}{Z^{2}}}\right)  =\left(
{\frac{P_{M}^{B}}{Z^{2}}}\right)  _{tr,lb}\tag{2.12}%
\end{equation}
where%

\begin{equation}
{\frac{1}{Z}}\left(  P_{ef}^{a}\right)  _{tr,lb}\text{ \ \ and \ \ }{\frac
{1}{Z}}\left(  \bar{P}_{ef}^{a}\right)  _{tr,lb}\tag{2.13}%
\end{equation}
are the independent probabilities of \textit{a }and \textit{\={a} }states
determined according to formulae (4.10 and 4.18 of [2,3]), while probabilities
$\left(  P_{ef}^{b}/Z\right)  _{tr,lb}$ and $\left(  \bar{P}_{ef}%
^{b}/Z\right)  _{tr,lb}$ are determined according to formulae (4.11 and 4.19
of [2,3]);

$Z\;$ is the sum of probabilities of all types of quasiparticles states -
eq.(4.2 of [2, 3]).

The probabilities of molecules being involved in the spatially independent
translational and librational macrodeformons are expressed as the products
(2.11) and (2.12):%

\begin{equation}
\left(
\begin{array}
[c]{c}%
P_{D}^{M}%
\end{array}
\right)  _{tr,lb}^{\text{ind}}=\left[
\begin{array}
[c]{c}%
\left(
\begin{array}
[c]{c}%
P_{M}^{A}%
\end{array}
\right)  ^{\text{ind}}\cdot\left(
\begin{array}
[c]{c}%
P_{M}^{B}%
\end{array}
\right)  ^{\text{ind}}%
\end{array}
\right]  _{tr,lb}={\frac{P_{D}^{M}}{Z^{4}}}\tag{2.14}%
\end{equation}
Formulae (2.11) and (2.12) may be considered as the probabilities of
space-independent but coherent macroeffectons in A and B states, respectively.

For probabilities of space-independent supereffectons in $A^{*}$ and $B^{*}$
states we have:%

\begin{equation}
\left(
\begin{array}
[c]{c}%
P_{S}^{A^{*}}%
\end{array}
\right)  ^{\text{ind}}=\left(
\begin{array}
[c]{c}%
P_{M}^{A}%
\end{array}
\right)  _{tr}^{\text{ind}}\cdot\left(
\begin{array}
[c]{c}%
P_{M}^{A}%
\end{array}
\right)  _{lb}^{\text{ind}}={\frac{P_{S}^{A^{*}}}{Z^{4}}}\tag{2.15}%
\end{equation}%
\begin{equation}
\left(
\begin{array}
[c]{c}%
P_{S}^{B^{*}}%
\end{array}
\right)  ^{\text{ind}}=\left(
\begin{array}
[c]{c}%
P_{M}^{B}%
\end{array}
\right)  _{tr}^{\text{ind}}\cdot\left(
\begin{array}
[c]{c}%
P_{M}^{B}%
\end{array}
\right)  _{tr}^{\text{ind}}={\frac{P_{S}^{b^{*}}}{Z^{4}}}\tag{2.15a}%
\end{equation}
\medskip In a similar way we get from (2.14) the probabilities of spatially
independent superdeformons:%

\begin{equation}
\left(
\begin{array}
[c]{c}%
P_{S}^{D^{*}}%
\end{array}
\right)  ^{\text{ind}}=\left(
\begin{array}
[c]{c}%
P_{M}^{D}%
\end{array}
\right)  _{tr}\cdot\left(
\begin{array}
[c]{c}%
P_{M}^{D}%
\end{array}
\right)  _{lb}={\frac{P_{S}^{D^{*}}}{Z^{4}}}\tag{2.16}%
\end{equation}
The concentrations of molecules, the states of which markedly differ from the
equilibrium one and which cause light scattering in composition of spatially
independent macroeffectons and macrodeformons, are equal to:%

\begin{equation}
\left[  N_{M}^{A}={\frac{N_{0}}{Z^{2}V_{0}}}\left(
\begin{array}
[c]{c}%
P_{M}^{A}%
\end{array}
\right)  \right]  _{tr,lb};\;\;\;\;\left[  N_{M}^{B}={\frac{N_{0}}{Z^{2}V_{0}%
}}\left(
\begin{array}
[c]{c}%
P_{M}^{B}%
\end{array}
\right)  \right]  _{tr,lb}\tag{2.17}%
\end{equation}%
\[
\left[  N_{M}^{D}={\frac{N_{0}}{Z^{4}V_{0}}}\left(
\begin{array}
[c]{c}%
P_{M}^{D}%
\end{array}
\right)  \right]  _{tr,lb}
\]
The concentrations of molecules, involved in a-convertons, b- convertons and
Macroconvertons or c-Macrotransitons (see Introduction) are correspondingly:%

\begin{equation}
N_{M}^{ac}={\frac{N_{0}}{Z^{2}V_{0}}}P_{ac};\qquad N_{M}^{bc}={\frac{N_{0}%
}{Z^{2}V_{0}}}P_{bc};\qquad N_{M}^{C}={\frac{N_{0}}{Z^{4}V_{0}}}P_{\text{cMt}%
}\tag{2.18}%
\end{equation}
The probabilities of convertons-related excitations are the same as used in
Chapter 4 of book [1].

The concentration of molecules, participating in the independent
supereffectons and superdeformons:%

\begin{equation}
N_{M}^{A^{*}}={\frac{N_{0}}{Z^{4}V_{0}}}P_{s}^{A^{*}};\qquad N_{M}^{B^{*}%
}={\frac{N_{0}}{Z^{4}V_{0}}}P_{S}^{B^{*}}\tag{2.19}%
\end{equation}%
\begin{equation}
N_{M}^{D^{*}}={\frac{N_{0}}{Z^{8}V_{0}}}P_{S}^{D^{*}}\tag{2.20}%
\end{equation}
where$\;N_{0}$ and $V_{0}$ are the Avogadro number and the molar volume of the matter.

Substituting (2.17 - 2.20) into well known \textbf{Raleigh formula for
scattering coefficient, measured at the right angle between incident and
scattered beams:}
\begin{equation}
R=\frac{I}{I_{0}}\frac{r^{2}}{V}=\frac{8\pi^{4}}{\lambda^{4}}\alpha
^{2}n_{\text{M}}\,\,\,(cm^{-1})\tag{2.20a}%
\end{equation}
we obtain the values of contributions from different states of quasiparticles
to the resulting scattering coefficient:%

\begin{equation}
\left(  R_{A}^{M}\right)  _{tr,lb}={\frac{8\pi^{4}}{\lambda^{4}}}%
{\frac{(\alpha^{*})^{2}}{Z^{2}}}{\frac{N_{0}}{V_{0}}}\left(
\begin{array}
[c]{c}%
P_{M}^{A}%
\end{array}
\right)  _{tr,lb};\;\;\,R_{A}^{s}={\frac{8\pi^{4}}{\lambda^{4}}}{\frac
{(\alpha^{*})^{2}}{Z^{4}}}{\frac{N_{0}}{V_{0}}}P_{s}^{A^{*}}\tag{2.21}%
\end{equation}%
\begin{equation}
\left(  R_{B}^{M}\right)  _{tr,lb}={\frac{8\pi^{4}}{\lambda^{4}}}%
{\frac{(\alpha^{*})^{2}}{Z^{2}}}{\frac{N_{0}}{V_{0}}}\left(
\begin{array}
[c]{c}%
P_{M}^{B}%
\end{array}
\right)  _{tr,lb};\;\;\,R_{B}^{s}={\frac{8\pi^{4}}{\lambda^{4}}}{\frac
{(\alpha^{*})^{2}}{Z^{4}}}{\frac{N_{0}}{V_{0}}}P_{s}^{B^{*}}\tag{2.22}%
\end{equation}%
\begin{equation}
\left(  R_{D}^{M}\right)  _{tr,lb}={\frac{8\pi^{4}}{\lambda^{4}}}%
{\frac{(\alpha^{*})^{2}}{Z^{2}}}{\frac{N_{0}}{V_{0}}}\left(
\begin{array}
[c]{c}%
P_{M}^{D}%
\end{array}
\right)  _{tr,lb};\;\;\,R_{D}^{s}={\frac{8\pi^{4}}{\lambda^{4}}}{\frac
{(\alpha^{*})^{2}}{Z^{4}}}{\frac{N_{0}}{V_{0}}}P_{s}^{D^{*}}\tag{2.23}%
\end{equation}
\medskip The contributions of excitations, related to $[tr/lb]$ convertons are:%

\[
R_{ac}={\frac{8\pi^{4}}{\lambda^{4}}}{\frac{(\alpha^{*})^{2}}{Z^{2}}}%
{\ \frac{N_{0}}{V_{0}}}R_{bc}={\frac{8\pi^{4}}{\lambda^{4}}}{\frac{(\alpha
^{*})^{2}}{Z^{2}}}{\frac{N_{0}}{V_{0}}}P_{bc}
\]
\[
R_{\text{abc}}={\frac{8\pi^{4}}{\lambda^{4}}}{\frac{(\alpha^{*})^{2}}{Z^{4}}%
}{\frac{N_{0}}{V_{0}}}P_{\text{cMt}}
\]
where: $\alpha^{*}$ is the acting polarizability determined by eq.(1.24) and (1.25).

The resulting coefficient of the isotropic scattering $(R_{\text{iso}})$ is
defined as the sum of contributions (2.21-2.23) and is subdivided into three
kinds of scattering: caused by translational quasiparticles, caused by
librational quasiparticles and by the mixed type of quasiparticles:
\begin{equation}
R_{\text{iso}}=[R_{A}^{M}+R_{B}^{M}+R_{D}^{M}]+[R_{A}^{M}+R_{B}^{M}+R_{D}%
^{M}]+[R_{ac}+R_{bc}+R_{\text{abc}}]+[R_{A}^{s}+R_{B}^{s}+R_{D}^{s}]\tag{2.24}%
\end{equation}

\thinspace\thinspace\ \ \ \ \ \ \ \ \ \ \ \ \ \ 

Total contributions, related to convertons and superexcitations are correspondingly:%

\[
R_{C}=R_{ac}+R_{bc}+R_{\text{abc\ \ }}\text{ and\ \ \ }R_{S}=R_{A}^{s}%
+R_{B}^{s}+R_{D}^{s}
\]

The polarizability of anisotropic molecules having no cubic symmetry is a
tensor. In this case, total scattering (R) consists of scattering at density
fluctuations $(R_{\text{iso}})$ and scattering at fluctuations of the
anisotropy $\left(  R_{\text{an}}={\frac{13\Delta}{6-7\Delta}}R_{\text{iso}%
}\right)  $:%

\begin{equation}
R=R_{\text{iso}}+{\frac{13\Delta}{6-7\Delta}}R_{\text{iso}}=R_{\text{iso}%
}{\ \frac{6+6\Delta}{6-7\Delta}}=R_{\text{iso}}K\tag{2.25}%
\end{equation}

where R$_{\text{iso}}$ corresponds to $eq.(2.24);\;\Delta$ is the
depolarization coefficient.

The factor: $\left(  {\frac{6+6\Delta}{6-7\Delta}}\right)  =K$ \ \ was
obtained by Cabanne and is called after him. In the case of isotropic
molecules when\thinspace$\;\Delta=0$, the Cabanne factor is equal to 1.

The depolarization coefficient ($\Delta$) could be determined experimentally
as the ratio:%

\begin{equation}
\Delta=I_{x}/I_{z},\tag{2.26}%
\end{equation}
where$\;I_{x}$ and $I_{z}$ are two polarized components of the beam scattered
at right angle with respect to each other in which the electric vector is
directed parallel and perpendicular to the incident beam, respectively. For
example, in water $\Delta=0.09\;($Vuks, 1977).

According to the proposed theory of light scattering in liquids the central
unshifted (like in gases) component of the Brillouin scattering spectrum, is
caused by fluctuations of concentration and self-diffusion of molecules,
participating in the convertons, macrodeformons (tr and lib) and
superdeformons. The scattering coefficients of the central line
$(R_{\text{centr}})$ and side lines $(2R_{\text{side}})$ in transparent
condensed matter, as follows from (2.24) and (2.25), are equal correspondingly to:%

\begin{equation}
R_{\text{cent}}=K\left[  \left(
\begin{array}
[c]{c}%
R_{D}^{M}%
\end{array}
\right)  _{tr}+\left(
\begin{array}
[c]{c}%
R_{D}^{M}%
\end{array}
\right)  _{lb}\right]  +K(R_{C}+R_{S})\tag{2.27}%
\end{equation}

and%

\begin{equation}
2R_{\text{side}}=\left(
\begin{array}
[c]{c}%
R_{A}^{M}+R_{B}^{M}%
\end{array}
\right)  _{tr}+\left(
\begin{array}
[c]{c}%
R_{A}^{M}+R_{A}^{M}%
\end{array}
\right)  _{lb}\tag{2.27a}%
\end{equation}
\smallskip

where $K$ is the Cabanne factor.

\smallskip

\textbf{The total coefficient of light scattering is:}%

\begin{equation}
R_{t}=R_{\text{cent}}+2R_{\text{side}}\tag{2.28}%
\end{equation}

In accordance with our model the fluctuations of anisotropy (Cabanne factor)
should be taken into account for calculations of the central component only.
The orientations of molecules in composition of A and B states of
Macroeffectons are correlated and their coherent oscillations are not
accompanied by fluctuations of anisotropy of polarizability (see Fig.2.1).

The probabilities of the convertons, macrodeformons and superdeformons
excitations (eqs.2.14,\ 4.16,\ 4.27 in [1]) are much lower in crystals than in
liquids and hence, the central line in the Brillouin spectra of crystals is
not usually observed.

The lateral lines in Brillouin spectra are caused by the scattering on the
molecules forming (A) and (B) states of spatially independent macroeffectons,
as it was mentioned above.

The polarizabilities of the molecules forming the independent macroeffectons,
synchronized in (A)$_{tr,lb}$ and $(B)_{tr,lb}$ states and dielectric
properties of these states, differ from each other and from that of transition
states (macrodeformons). Such short-living states should be considered as the
non equilibrium ones.

In fact we must keep in mind, that static polarizabilities in the more stable
ground A state of the macroeffectons are higher than in B state, because the
energy of long-term Van der Waals interaction between molecules of the A state
is bigger than that of B-state.

If this difference may be attributed mainly to the difference in the
long-therm dispersion interaction, then from (1.33) we obtain:%

\begin{equation}
E_{B}-E_{A}=V_{B}-V_{A}=-{\frac{3}{2}}{\frac{E_{0}}{r^{6}}}\left(
\begin{array}
[c]{c}%
\alpha_{B}^{2}-\alpha_{A}^{2}%
\end{array}
\right) \tag{2.29}%
\end{equation}
where polarizability of molecules in A-state is higher, than that in B-state:%

\[
\alpha_{A}^{2}>\left[
\begin{array}
[c]{c}%
\left(
\begin{array}
[c]{c}%
\alpha^{*}%
\end{array}
\right)  ^{2}\approx\alpha_{D}^{2}%
\end{array}
\right]  >\alpha_{B}^{2}
\]
\medskip The kinetic energy and dimensions of ''acoustic'' and ''optic''
states of macroeffectons are the same: $T_{\text{kin}}^{A}=T_{\text{kin}}^{B}$.

In our present calculations of light scattering we ignore this difference
(2.29) between polarizabilities of molecules in A and B states.

But it can be taken into account if we assume, that polarizabilities in (A)
and (a), (B) and (b) states of primary effectons are like:%

\[
\alpha_{A}\simeq\alpha_{a}\simeq\alpha^{*};\;\;\alpha_{B}\simeq\alpha_{b}
\]
and the difference between the potential energy of (a) and (b) states is
determined mainly by dispersion interaction (eq.2.28).

Experimental resulting polarizability ($\alpha^{*} \simeq\alpha_{a}$) can be
expressed as:%

\begin{equation}
\alpha_{a}=f_{a}\alpha_{a}+f_{b}\alpha_{b}+f_{t}\alpha\tag{2.29a}%
\end{equation}
where $\alpha_{t}\simeq\alpha$ is polarizability of molecules in the gas state
(or transition state);%

\begin{align*}
f_{a}  & ={\frac{P_{a}}{P_{a}+P_{b}+P_{t}}};\;\;\;f_{b}={\frac{P_{b}}%
{P_{a}+P_{b}+P_{t}}}\text{;}\\
\text{ and\ \ \ }f_{t}  & =f_{d}={\frac{P_{t}}{P_{a}+P_{b}+P_{t}}}%
\end{align*}
are the fractions of (a), (b) and transition (t) states (equal to 2.66) as far
$P_{t}=P_{d}=P_{a}\cdot P_{b}$.

On the other hand from (1.33) at\ $r=const$ \ we have:%

\[
\Delta V_{\text{dis}}^{b\rightarrow a}=-{\frac{3}{4}}{\frac{(2\alpha
\Delta\alpha)}{r^{6}}}\cdot I_{0}\;\;\;\;(r_{a}=r_{b};\;\;I_{0}^{a}\simeq
I_{0}^{b})\text{ and }
\]
\begin{equation}
{\frac{\Delta V_{\text{dis}}^{b\rightarrow a}}{V^{b}}}={\frac{h\nu_{p}}%
{h\nu_{b}}}={\frac{\Delta\alpha_{a}}{\alpha}\;}\text{ or \ }\Delta\alpha
_{a}=\alpha_{a}{\ \frac{\nu_{p}}{\nu_{b}}}\tag{2.29b}%
\end{equation}%

\[
\alpha_{b}=\alpha_{a}-\Delta\alpha_{a}=\alpha_{a}(1-\nu_{p}/\nu_{b})
\]

where:\thinspace$\Delta\alpha_{a}$ is a change of each molecule polarizability
as a result of the primary effecton energy changing: $\,E_{b}\rightarrow
E_{a}+h\nu_{p}$ with photon radiation; $\nu_{b}$ is a frequency of primary
effecton in (b)- state (eq.2.28).

Combining (2.29) and (2.29b) we derive for $\alpha_{a}$ and $\alpha_{b}$ of
the molecules composing primary translational or librational effectons:%

\begin{equation}
\alpha_{a}={\frac{f_{t}\alpha}{1-\left(  f_{a}+f_{b}+f_{b}\cdot{\frac{\nu_{p}%
}{\nu_{b}}}\right)  }}\tag{2.30}%
\end{equation}%
\begin{equation}
\alpha_{b}=\alpha_{a}\left(  1-{\frac{\nu_{p}}{\nu_{b}}}\right) \tag{2.30a}%
\end{equation}
The calculations by means of (2.30) are approximate in the framework of our
assumptions mentioned above. But they correctly reflect the tendencies of
$\alpha_{a}$ and $\alpha_{b}$ changes with temperature.

The ratio of intensities or scattering coefficients for the central component
to the lateral ones previously was described by Landau- Plachek formula
(2.10). According to our mesoscopic theory this ratio can be calculated in
another way leading from (2.27) and (2.28):%

\begin{equation}
{\frac{I_{\text{centr}}}{2I_{M-B}}}={\frac{R_{\text{cent}}}{2R_{\text{side}}}%
}\tag{2.30b}%
\end{equation}

Combining (2.30) and Landau- Plachek formula (2.10) it is possible to
calculate the ratio $(\beta_{T}/\beta_{S})$ and $(C_{P}/C_{V})$ using our
mesoscopic theory of light scattering.

\smallskip

\begin{center}
{\large 2.4. Factors that determine the Brillouin line width}
\end{center}

\smallskip

The known equation for Brillouin shift is (see 2.7):%

\begin{equation}
\Delta\nu_{M-B}=\nu_{0}=2{\frac{v_{s}}{\lambda}}n\sin(\theta/2)\tag{2.31}%
\end{equation}
where: $v_{s}$ is the hypersonic velocity; $\lambda$ \thinspace\thinspace is
the wavelength of incident light, $n$ is the refraction index of matter, and
$\theta$ - scattering angle.

The deviation from $\nu_{0}$ that determines the Brillouin side line half
width \textit{may be expressed as the result of fluctuations of sound velocity
}$v_{s}$ \textit{and }$\mathit{n}$\textit{\ }related to A and B states of tr
and lib macroeffectons:%

\begin{equation}
{\frac{\Delta\nu_{0}}{\nu_{0}}}=\left(  {\frac{\Delta v_{s}}{v_{s}}}%
+{\frac{\Delta n}{n}}\right) \tag{2.32}%
\end{equation}

$\Delta\nu_{0}$ is the most probable side line width, i.e. the true half width
of Brillouin line. It can be expressed as:%

\[
\Delta\nu_{0}=\Delta\nu_{\exp}-F\Delta\nu_{\text{inc}}
\]
where $\Delta\nu_{\exp}$ is the half width of the experimental line,
$\Delta\nu_{\text{inc}}$ - the half width of the incident line, $F$ - the
coefficient that takes into account apparatus effects.

Let us analyze the first and the second terms in the right part of (2.32) separately.

The\thinspace$v_{s}$ squared is equal to the ratio of the compressibility
modulus ($K$) and density $(\rho)$:%

\begin{equation}
v_{s}^{2}=K^{2}/\rho\tag{2.33}%
\end{equation}
Consequently, from (2.33) we have:%

\begin{equation}
{\frac{\Delta v_{s}}{v_{s}}}={\frac{1}{2}}\left(  {\frac{\Delta K}{K}}%
-{\frac{\Delta\rho}{\rho}}\right) \tag{2.34}%
\end{equation}
In the case of independent fluctuations of K and $\rho:$:%

\begin{equation}
{\frac{\Delta v_{s}}{v_{s}}}={\frac{1}{2}}\left(  \left|  {\frac{\Delta K}{K}%
}\right|  -\left|  {\frac{\Delta\rho}{\rho}}\right|  \right) \tag{2.35}%
\end{equation}
From our equation (1.14) we obtain for refraction index:%

\begin{equation}
n^{2}=\left(  1-{\frac{4}{3}}N\alpha^{*}\right)  ^{-1},\tag{2.36}%
\end{equation}
where $N=N_{0}/V_{0}$ is the concentration of molecules.

From (2.36) we can derive:%

\begin{equation}
{\frac{\Delta n}{n}}={\frac{1}{2}}\left(  n^{2}-1\right)  \left(
{\frac{\Delta\alpha^{*}}{\alpha^{*}}}+{\frac{\Delta N}{N}}\right) \tag{2.37}%
\end{equation}
where:%

\begin{equation}
(\Delta N/N)=(\Delta\rho/\rho)\tag{2.38}%
\end{equation}
and%

\begin{equation}
\left(  {\frac{\Delta\alpha^{*}}{\alpha^{*}}}\right)  \simeq\left(
{\frac{\Delta K}{K}}\right) \tag{2.39}%
\end{equation}
we can assume eq.(2.39) as far both parameters: polarizability $(\alpha^{*}) $
and compressibility models (K) are related with the potential energy of
intermolecular interaction.

For the other hand one can suppose that the following relation is true:%

\begin{equation}
{\frac{\Delta\alpha^{*}}{\alpha^{*}}}\simeq{\frac{\mid\bar{E}_{ef}^{a}%
-3kT\mid}{3kT}}={\frac{\Delta K}{K}}\tag{2.40}%
\end{equation}

where: $%
\begin{array}
[c]{l}%
\bar E_{ef}^{a}%
\end{array}
$ is the energy of the secondary effectons in (\=a) state;\ $E_{0}=3kT\;$ is
the energy of an ''ideal'' quasiparticle as a superposition of 3D standing waves.

The density fluctuations can be estimated as a result of the free volume
$(v_{f})$ fluctuations (see 2.45):%

\begin{equation}
\left(  {\frac{\Delta v_{f}}{v_{f}}}\right)  _{tr,lb}={\frac{1}{Z}}\left(
P_{D}^{M}\right)  _{tr,lb}\simeq\left(  \Delta N/N\right)  _{tr,lb}\tag{2.41}%
\end{equation}
Now, putting (2.40) and (2.41) into (2.37) and (2.34) and then into (2.32), we
obtain the semiempirical formulae for the Brillouin line half width calculation:%

\begin{equation}
{\frac{\Delta\nu_{f}}{\nu_{f}}}\simeq{\frac{n^{2}}2}\left[  {\frac{\mid\bar
E_{ef}^{a}-3kT\mid}{3kT}}+{\frac1Z}\left(
\begin{array}
[c]{c}%
P_{D}^{M}%
\end{array}
\right)  \right]  _{tr,lb}\tag{9.42}%
\end{equation}

Brillouin line intensity depends on the half-width $\Delta\nu$ of the line in
following ways:

for a Gaussian line shape:%

\begin{equation}
I(\nu)=I_{0}^{\max}\exp\left[  -0.693\left(  {\frac{\nu-\nu_{0}}{{\frac{1}{2}%
}\Delta\nu_{0}}}\right)  ^{2}\right]  ;\tag{2.43}%
\end{equation}
for a Lorenzian line shape:%

\begin{equation}
I(\nu)={\frac{I_{0}^{\max}}{1+\left[
\begin{array}
[c]{c}%
\left(  \nu-\nu_{0}\right)  /{\frac{1}{2}}\Delta\nu_{0}%
\end{array}
\right]  ^{2}}}\tag{2.44}%
\end{equation}
\medskip\textit{The traditional theory of Brillouin line shape }gives a
possibility for calculation of $\Delta\nu_{0}$ taking into account the elastic
(acoustic) wave dissipation.

The fading out of acoustic wave amplitude may be expressed as:%

\begin{equation}
A=A_{0}e^{-\alpha x}\text{ \ \ or\ \ }A=A_{0}e^{-\alpha v_{s}}\tag{2.45}%
\end{equation}
where $\alpha$ is the extinction coefficient;\ $\;x=v_{s}t$ - the distance
from the source of waves;\ $v_{s}$ and $\;t$ - sound velocity and time, correspondingly.

\medskip

The hydrodynamic theory of sound propagation in liquids leads to the following
expression for the extinction coefficient:%

\begin{equation}
\alpha=\alpha_{s}+\alpha_{b}={\frac{\Omega^{2}}{2\rho v_{s}^{3}}}\left(
{\frac{4}{3}}\eta_{s}+\eta_{b}\right) \tag{2.46}%
\end{equation}
where: $\alpha_{s}\;$and $\alpha_{b}$ are contributions to $\alpha$, related
to share viscosity ($\eta_{s}$) and bulk viscosity $(\eta_{b})$, respectively;
$\Omega=2\pi f$ \ is the angular frequency of acoustic waves.

When the side lines in Brillouin spectra broaden slightly, the following
relation between their intensity (I) and shift $(\Delta\omega= \mid\omega-
\omega_{0}\mid)$ from frequency $\omega_{0}$, corresponding to maximum
intensity $(I = I_{0})$ of side line is correct:%

\begin{equation}
I={\frac{I_{0}}{1+\left(  {\frac{\omega-\omega_{0}}{a}}\right)  }},\tag{2.47}%
\end{equation}
where:%

\[
a=\alpha v_{s}.
\]
One can see from (2.46) that at $\;I(\omega)=I_{0}/2$, the half width:%

\begin{equation}
\Delta\omega_{1/2}=2\pi\Delta\nu_{1/2}=\alpha v_{s}\text{ \ \ and\ \thinspace
\thinspace\ }\Delta\nu_{1/2}={\frac{1}{2}}\pi\alpha v_{s}\tag{2.48}%
\end{equation}
It will be shown in Chapter 12 how one can calculate the values of $\eta_{s} $
and consequently $\alpha_{s}$ on the basis of the mesoscopic theory of viscosity.

\smallskip

\begin{center}
{\large 2.5. Quantitative verification of mesoscopic theory of}

{\large Brillouin scattering}
\end{center}

\textbf{\smallskip}

\textbf{The calculations made according to the formula (2.21 - 2.27) are
presented in Fig.2.1-2.7. The proposed theory of scattering in liquids, based
on our hierarchic concept, is more adequate than the traditional Einstein,
Mandelschtamm-Brillouin, Landau-Plachek theories based on classical
thermodynamics. It describes experimental temperature dependencies and the
}$I_{\text{centr}}/2I_{M-B}$\textbf{\ ratio for water very well (Fig.2.3).}

The calculations are made for the wavelength of incident light: $\lambda
_{ph}=546.1nm=5.461\cdot10^{-5}cm$. The experimental temperature dependence
for the refraction index (n) at this wavelength was taken from the Frontas'ev
and Schreiber paper (1965). The rest of data for calculating of various light
scattering parameters of water (density the location of translational and
librational bands in the oscillatory spectra) are identical to those used
above in Chapter 6.

\begin{center}%
\begin{center}
\includegraphics[
height=2.6152in,
width=4.8672in
]%
{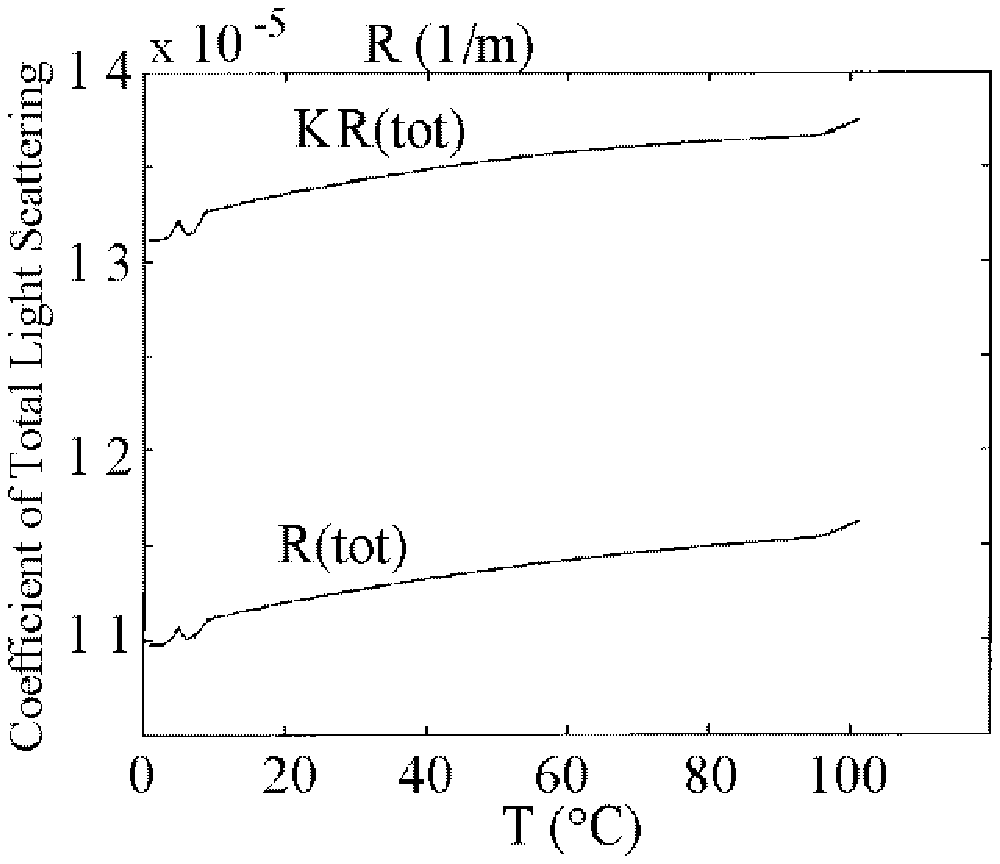}%
\end{center}
\end{center}

\begin{quotation}
\textbf{Fig.~2.1. }Theoretical temperature dependencies of the total
scattering coefficient for water without taking into account the anisotropy of
water molecules polarizability fluctuations in the volume of macroeffectons,
responsible for side lines: $\left[  R(tot)\right]  $ - eq.(2.27a; 2.28) and
taking them into account: $\left[  KR(tot)\right]  $, where $K\,$ is the
Cabanne factor (eq.2.25).\medskip%
\begin{center}
\includegraphics[
height=2.738in,
width=5.0912in
]%
{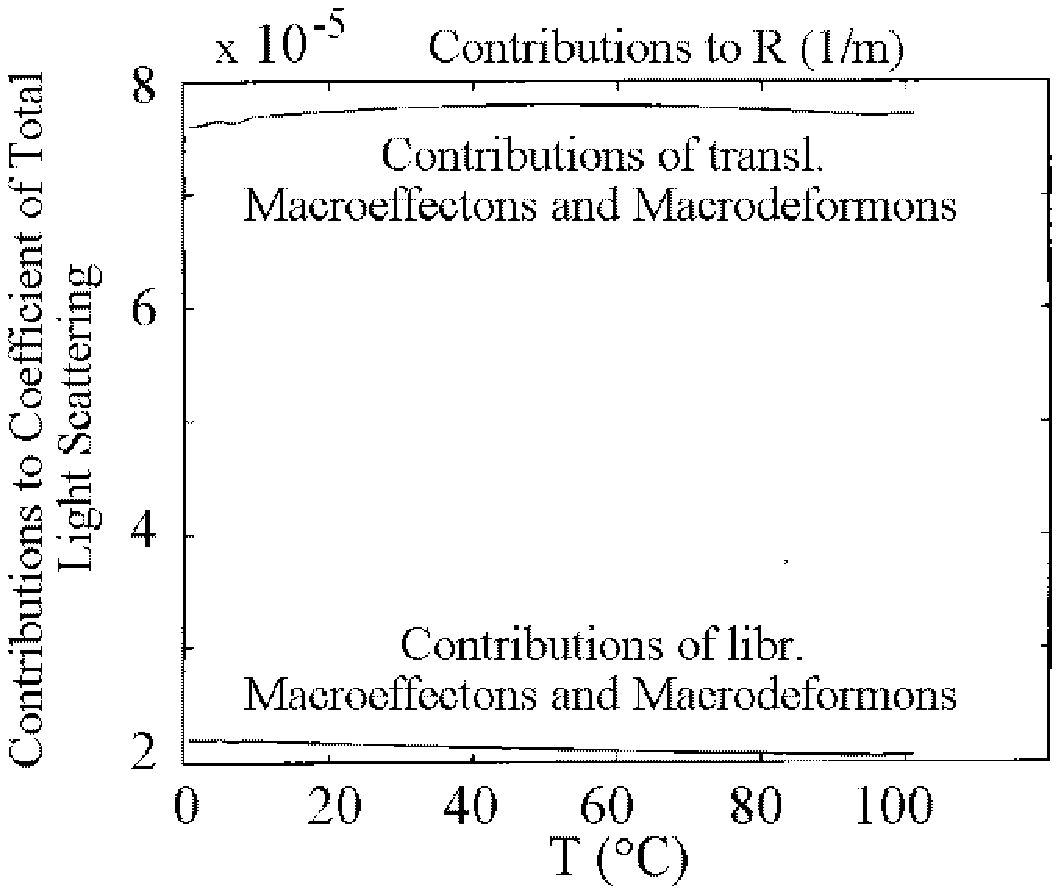}%
\end{center}

\textbf{Fig.~2.2. }Theoretical temperature dependencies of contributions to
the total coefficient of total light scattering (R) caused by translational
and librational macroeffectons and macrodeformons (without taking into account
fluctuations of anisotropy).\medskip
\end{quotation}

\begin{center}%
\begin{center}
\includegraphics[
height=2.8297in,
width=5.0341in
]%
{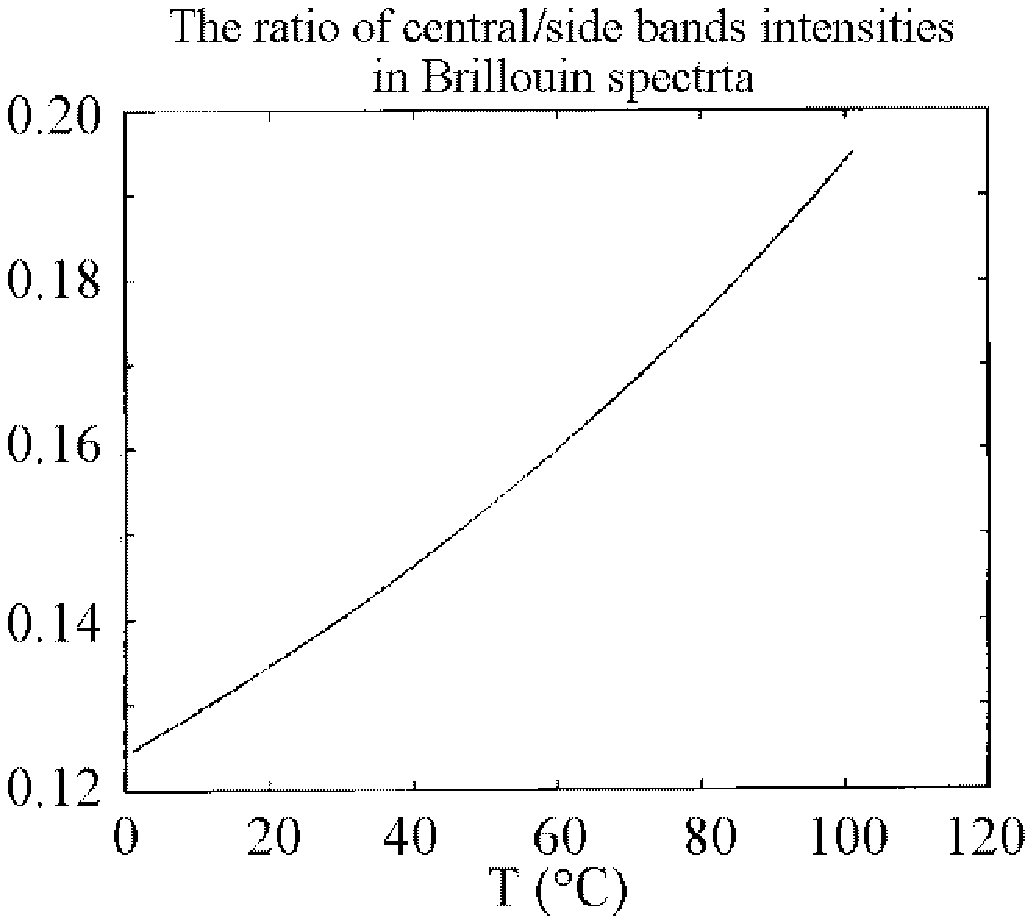}%
\end{center}
\end{center}

\begin{quotation}
\textbf{Fig.~2.3. }Theoretical temperature dependencies of central to side
bands intensities ratio in Brillouin spectra (eq.2.30).
\end{quotation}

\medskip

\ Mesoscopic theory of light scattering can be used to verify the correctness
of our formula for refraction index of condensed matter we got from our theory
(eq. 1.14):
\begin{equation}
\frac{n^{2}-1}{n^{2}}=\frac{4}{3}\pi{\frac{N_{0}}{V_{0}}}\alpha^{*}\tag{2.48a}%
\end{equation}

and to compare the results of its using with that of the Lorentz-Lorenz formula:%

\begin{equation}
{\frac{n^{2}-1}{n^{2}+1}}={\frac{4}{3}}\pi{\frac{N_{0}}{V_{0}}}\alpha
\tag{2.49}%
\end{equation}
\ \ 

\ From formula (2.48a) the resulting or effective molecular polarizability
squared $(\alpha^{*})^{2}$ used in eq.(2.21-2.23) is:%

\begin{equation}
\left(  \alpha^{*}\right)  ^{2}=\left[  {\frac{(n^{2}-1)/n^{2}}{(4/3)\pi
(N_{0}/V_{0})}}\right]  ^{2}\tag{2.50}%
\end{equation}
On the other hand, from the Lorentz-Lorenz formula (2.49) we have another
value of polarizability:%

\begin{equation}
\alpha^{2}=\left[  {\frac{(n^{2}-1)/(n^{2}+2)}{(4/3)\pi(N_{0}/V_{0})}}\right]
^{2}\tag{2.51}%
\end{equation}

It is evident that the light scattering coefficients (eq.2.28), calculated
using (2.50) and (2.51) taking refraction index: $n=1.33\,\,\,$should differ
more than four times as far:%

\begin{equation}
{\frac{R(\alpha^{*})}{R(\alpha)}}={\frac{(\alpha^{*})^{2}}{(\alpha)^{2}}%
}={\frac{(n^{2}-1)/n^{2}}{(n^{2}-1)/(n^{2}+2)}}=\left(  {\frac{n^{2}+2}{n^{2}%
}}\right)  ^{2}=4.56\tag{2.52}%
\end{equation}

\ At $\;25^{0}$ and $\;\lambda_{ph}=546nm$ \ the theoretical magnitude of the
scattering coefficient for water, calculated from our formulae (2.28) is equal
(see Fig.2.1) to:%

\begin{equation}
R=11.2\cdot10^{-5}m^{-1}\tag{2.53}%
\end{equation}
This result of our theory coincides well with the most reliable experimental
value (Vuks, 1977):%

\[
R_{\exp}=10.8\cdot10^{-5}m^{-1}%
\]
Multiplication of the side bands contribution $(2R_{\text{side}})$ to Cabanne
factor increases the calculated total scattering to about 25\% and makes the
correspondence with experiment worse. This fact confirms our assumption that
fluctuations of anisotropy of polarizability in composition of A and B states
of macroeffectons should be ignored in light scattering evaluation due to
correlation of molecular dynamics in these states, in contrast to that of macrodeformons.

\begin{center}%
\begin{center}
\includegraphics[
height=2.9456in,
width=5.4751in
]%
{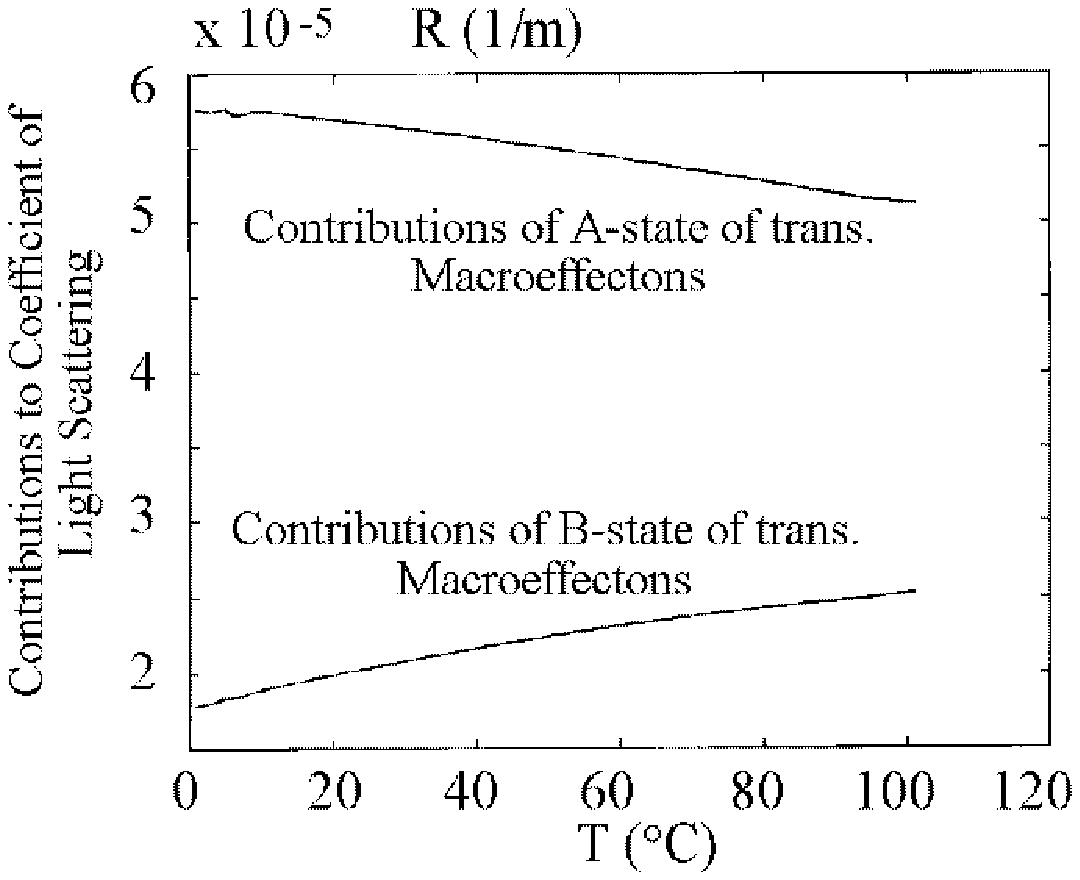}%
\end{center}
\end{center}

\begin{quotation}
\textbf{Fig.~2.4. }Theoretical temperature dependencies of the contributions
of A and B states of translational Macroeffectons to the total scattering
coefficient of water (see also Fig.2.2);\medskip
\end{quotation}

\begin{center}%
\begin{center}
\includegraphics[
height=2.514in,
width=4.6726in
]%
{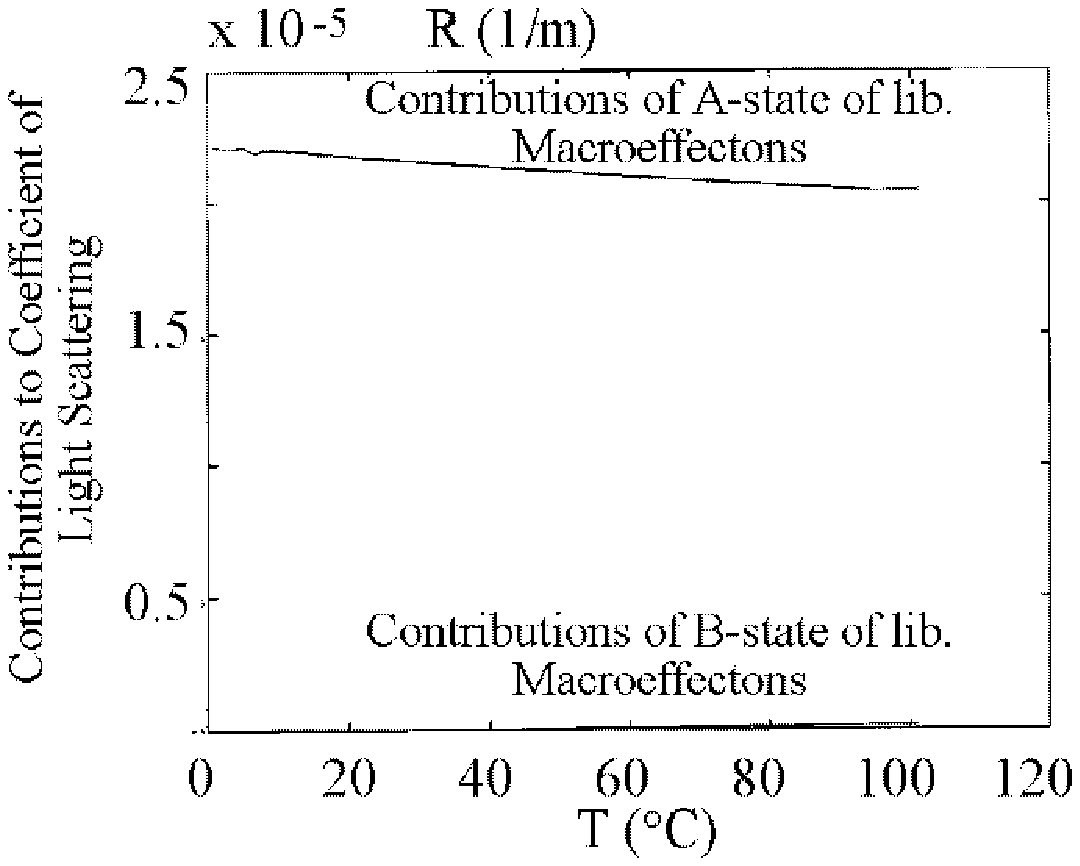}%
\end{center}
\end{center}

\begin{quotation}
\textbf{Fig.~2.5. }Theoretical temperature dependencies of the contributions
of the A and B states of librational Macroeffectons to the coefficient of
light scattering (R).
\end{quotation}

\medskip

It follows from the Fig.2.4 and 2.5 that the light scattering depends on
$(A\Leftrightarrow B)$ equilibrium of macroeffectons because $(R_{A})>(R_{B}%
)$, i.e. scattering on $A$ states is bigger than that on $B$ states.

\begin{center}%
\begin{center}
\includegraphics[
height=2.7873in,
width=5.1776in
]%
{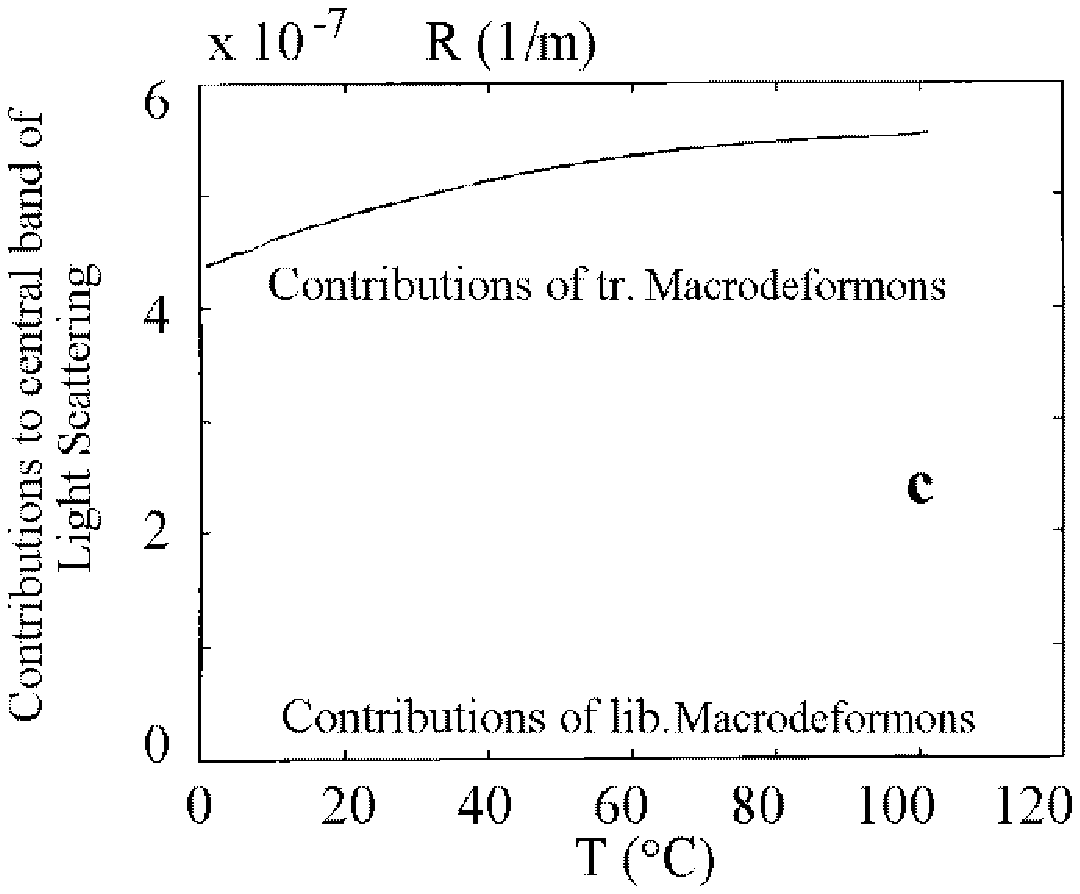}%
\end{center}
\end{center}

\begin{quotation}
\textbf{Fig.~2.6. }Theoretical temperature dependencies of the contributions
to light scattering (central component), related to translational
$(R_{D})_{tr}$ and librational $(R_{D})_{lb}$ macrodeformons.

\medskip
\end{quotation}

Comparing Figs. 2.1; 2.3, and 2.6 one can see that the main contribution to
central component of light scattering is determined by $[lb/tr]$ convertons
$R_{c}\;($see eq.2.27).

\begin{center}%
\begin{center}
\includegraphics[
height=2.9499in,
width=5.4803in
]%
{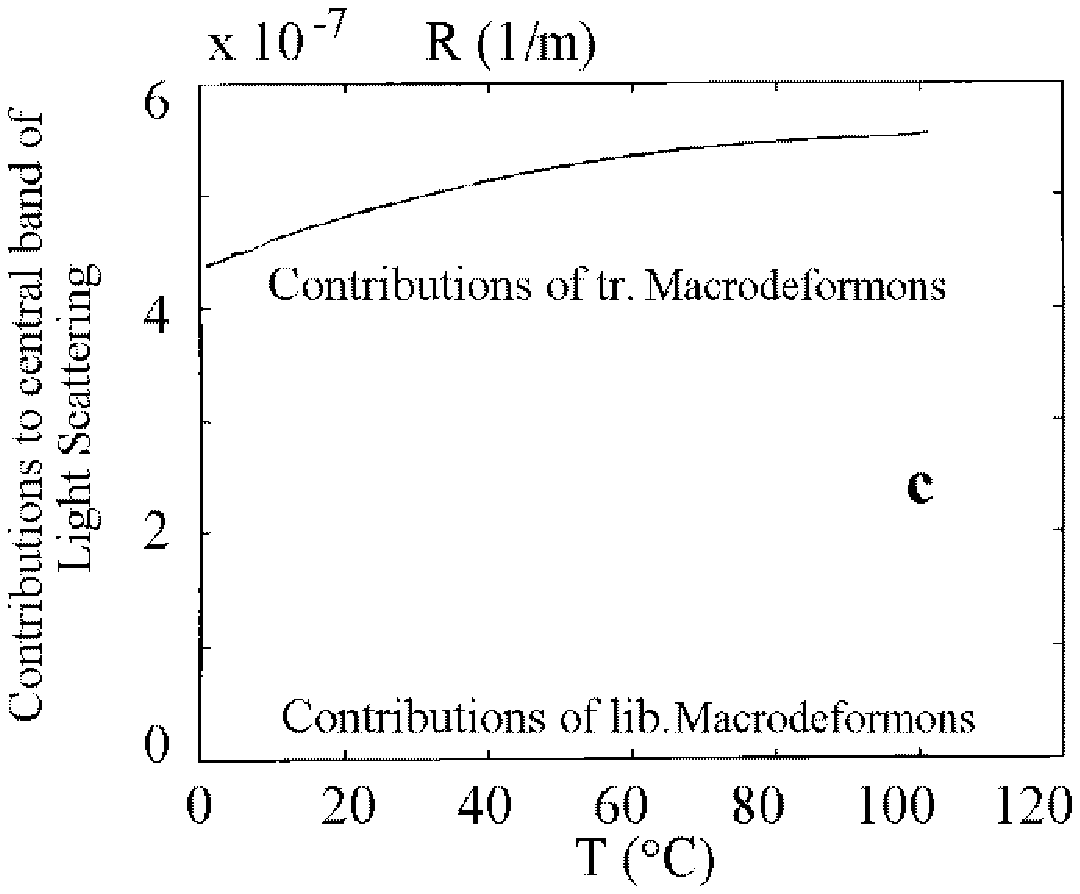}%
\end{center}
\end{center}

\begin{quotation}
\textbf{Fig.~2.7. }Theoretical temperature dependences for temperature
derivative $(dR/dT)$ of the total coefficient of light scattering of water.
\end{quotation}

\medskip

Nonmonotonic deviations of the dependencies $dR/dT\;($Fig.2.7) reflect the
nonmonotonic changes of the refraction index for water $n_{H_{2}O}(T)$, as
indicated by available experimental data (Frontas'ev and Schreiber, 1965). The
deviations of dependence $\,n_{H_{2}O}(t)$ from the monotonic way in
accordance with hierarchic theory, are a consequence of the nonmonotonic
change in the stability of water structure, i.e. nonlinear change of
$(A\Leftrightarrow B)_{tr,lb}$ equilibrium. Some possible reasons of such
equilibrium change were discussed in Chapter 6.

\textbf{It is clear from (2.52) that the calculations based on the
Lorentz-Lorentz formula (2.51) give scattering coefficient values of about 4.5
times smaller than experimental ones. It means that the true }$\alpha^{*}%
$\textbf{\ value can be calculated just on the basis of our mesoscopic theory
of light refraction (eq.2.50).}

The traditional Smolukhovsky-Einstein theory, valid for the integral light
scattering only (eq. 2.1), yield values in the range of $\;R=8.85\cdot
10^{-5}\,m^{-1}\;$ to\ $R=10.5\cdot10^{-5}\,m^{-1}\;[4$, 8$]$.

\textbf{All the results, discussed above, mean that our mesoscopic theory of
light scattering works better and is much more informative than the
conventional one.}

\medskip

\begin{center}

{\large 2.6. Light scattering in solutions}
\end{center}

\smallskip

If the guest molecules are dissolved in a liquid and their sizes are much less
than incident light wavelength, they do not radically alter the solvent
properties. For this case the described above mechanism of light scattering of
pure liquids does not changed qualitatively.

For such dilute solutions the scattering on the fluctuations of concentration
of dissolved molecules $(R_{c})$ is simply added to the scattering on the
density fluctuations of molecules of the host solvent (eq.2.28). Taking into
account the fluctuations of molecule polarizability anisotropy (see 2.25) the
total scattering coefficient of the solution $(R_{S})$ is:%

\begin{equation}
R_{S}=R_{t}+R_{c}\tag{2.54}%
\end{equation}

Eqs. (2.21 - 2.28) could be used for calculating $R_{t}$ until critical
concentrations $(C_{cr})$ of dissolved substance when it start to destroy the
solvent structure, so that the latter is no longer able to form primary
librational effectons. Perturbations of solvent structure will induce
low-frequency shift of librational bands in the oscillatory spectrum of the
solution until these bands totally disappear.

If the experiment is made with a two-component solution of liquids, soluble in
each other, e.g. water-alcohol, benzol-methanol etc., and the positions of
translational and librational bands of solution components are different, then
at the concentration of the dissolved substance: $C>C_{cr}$, the dissolved
substance and the solvent (the guest and host) can switch their roles. Then
the translational and librational bands pertinent to the guest subsystem start
to dominate. In this case, $R_{t}$ is to be calculated from the positions of
the new bands corresponding to the ''new'' host-solvent. The total ''melting''
of the primary librational ''host effectons'' and the appearance of the
dissolved substance ''guest effectons'' is \textit{like the second order phase
transition} and should be accompanied by a heat capacity jump. The like
experimental effects take place indeed [8].

According to our concept, the coefficient R$_{c}$ in eq.(2.54) is caused by
the fluctuations of concentration of dissolved molecules in the volume of
translational and librational macro- and superdeformons of the solvent. If the
destabilization of the solvent is expressed in the \textbf{low frequency
}shift of librational bands, then the coefficients $(R_{A}$ \thinspace
and\thinspace$R_{B})_{lb}$ increase (eq.2.21 and 2.22) with the probability of
macro-excitations.. The probabilities of convertons and macro- and
superdeformons and the central component of Brillouin spectra will increase
also. Therefore, the intensity of the total light scattering increases correspondingly.

The fluctuations of concentration of the solute molecules, in accordance with
our model, occur in the volumes of macrodeformons and superdeformons.
Consequently, the contribution of solute molecules in scattering $(R_{c}$
value in eq.2.54) can be expressed by formula, similar to (2.23), but
containing the molecule polarizability of the dissolved substance
(''guest$"),$ equal to $(\alpha_{g}^{*})^{2}$ instead of the molecule
polarizability $(\alpha^{*})$ of the solvent (''host''), and the molecular
concentration of the ''guest'' substance in the solution $(n_{g})$ instead of
the solvent molecule concentration $(n_{M}=N_{0}/V_{0})$. For this case
$R_{c}$ could be presented as a sum of the following contributions:%

\begin{equation}
(R_{c})_{tr,lb}={\frac{8\pi^{4}}{\lambda^{4}}}(\alpha_{g}^{*})^{2}n_{g}%
\cdot\left[  (P_{M}^{D})_{tr,lb}+P_{S}^{D^{*}}\right] \tag{2.55}%
\end{equation}%

\begin{equation}
R_{c}^{D^{*}}={\frac{8\pi^{4}}{\lambda^{4}}}(\alpha_{g}^{*})^{2}n_{g}%
\cdot(P_{S}^{D^{*}})\tag{2.55a}%
\end{equation}

The resulting scattering coefficient $(R_{e})$ on fluctuations of
concentration in (2.54) is equal to:%

\begin{equation}
R_{c}=(R_{c})_{tr}+(R_{c})_{lb}+R_{c}^{D^{*}}\tag{2.56}%
\end{equation}

If \textbf{several }substances are dissolved with concentrations lower than
$(C_{cr})$, then their $R_{c}$ are summed up additively.

Formulae (2.55) and (2.56) are valid also for the dilute solutions.

Eqs.(2.21-2.28) and (2.54-2.56) should, therefore, be used for calculating the
resulting coefficient of light scattering in solutions $(R_{S})$.

\textit{The traditional theory represents }the scattering coefficient at
fluctuations of concentration as (Vuks, 1977):%

\begin{equation}
R_{c}={\frac{\pi^{2}}{2\lambda^{4}}}\left(  {\frac{\partial\epsilon}{\partial
x}}\right)  ^{2}\Delta x^{2}v\tag{2.57}%
\end{equation}
where $(\partial\epsilon/\partial x)$ is the dielectric penetrability
derivative with respect to one of the components: $\Delta\bar{x}^{2}$ is the
fluctuations of concentration of guest molecules squared in the volume element
$\mathit{v}$.

The transformation of (2.57) on the basis of classical thermodynamics [8]
leads to the formula:%

\begin{equation}
R_{c}={\frac{\pi^{2}}{2\lambda^{4}N_{0}}}\left(  2n{\frac{\partial n}{\partial
x}}\right)  \left(  {\frac{9n^{2}}{(2n^{2}+1)(n^{2}+2)}}\right)  ^{2}%
x_{1}x_{2}V_{12}f,\tag{2.58}%
\end{equation}
where $N_{0}$ is the Avogadro number, $x_{1}$ and $x_{2}$ are the molar
fractions of the first and second components in the solution, $V_{12}$ is the
molar volume of the solution,\thinspace$\mathit{f\,\,}$ is the function of
fluctuations of concentration determined experimentally from the partial vapor
pressures of the first $(P_{1})$ and second $(P_{2})$ solution components [8]:%

\begin{equation}
{\frac{1}{f}}={\frac{x_{1}}{P_{1}}}{\frac{\partial P_{1}}{\partial x_{1}}%
}={\frac{x_{2}}{P_{2}}}{\frac{\partial P_{2}}{\partial x_{2}}}\tag{2.59}%
\end{equation}
In the case of ideal solutions%

\[
{\frac{\partial P_{1}}{\partial x_{1}}}={\frac{P_{1}}{x_{1}}};\;\;{\frac
{\partial P_{2}}{\partial x_{2}}}={\frac{P_{2}}{x_{2}}}\text{
;\ \ and\thinspace\ \ }f=1.
\]
\medskip\textbf{For application the mesoscopic theory of light scattering to
study of crystals, liquids and solutions, the following information is needed: }

\textbf{1.~Positions of translational and librational band maxima in
oscillatory spectra;}

\textbf{2.~Concentration of all types of molecules in solutions;}

\textbf{3.~Refraction index or polarizability in the acting field of each
component of solution at given temperature.}

\smallskip

\textbf{Application of our theory to quantitative analysis of transparent
liquids and solids yields much more information about properties of matter,
its mesoscopic and hierarchic dynamic structure than the traditional one}.

\bigskip

\bigskip

\begin{center}
{\Large 3. \ Mesoscopic theory of M\"{o}ssbauer effect\ }

\smallskip

{\large 3.1. General background}
\end{center}

\smallskip

When the atomic nucleus with mass (M) in the gas phase irradiates $\gamma
$-quantum with energy of%

\begin{equation}
E_{0}=h\nu_{0}=m_{p}c^{2}\tag{3.1}%
\end{equation}
where: $m_{p}$ is the effective photon mass, then according to the law of
impulse conservation, the nuclear acquires \textit{additional }velocity in the
opposite direction:%

\begin{equation}
v=-{\frac{E_{0}}{Mc}}\tag{3.2}%
\end{equation}
The corresponding additional kinetic energy%

\begin{equation}
E_{R}={\frac{Mv^{2}}{2}}={\frac{E_{0}^{2}}{2Mc^{2}}}\tag{3.3}%
\end{equation}
is termed \textbf{recoil energy. }

When an atom which irradiates $\gamma$-quantum is in composition of the solid
body, then \textbf{three situations }are possible:

\textbf{1.~The recoil energy of the atom is higher than the energy of atom -
lattice interaction. In this case, the atom irradiating }$\gamma
$\textbf{-quantum would be knocked out from its position in the lattice. That
leads to defects origination;}

\textbf{2.~Recoil energy is insufficient for the appreciable displacement of
an atom in the structure of the lattice, but is higher than the energy of
phonon, equal to energy of secondary transitons and phonons excitation. In
this case, recoil energy is spent for heating the lattice; }

\textbf{3.~Recoil energy is lower than the energy of primary transitons,
related to [emission/absorption] of IR translational and librational photons
}$(h\nu_{p})_{tr,lb}$\textbf{\ and phonons }$(h\nu_{ph})_{tr,lb}$\textbf{. In
that case, the probability (f) of }$\gamma$\textbf{-quantum irradiation
without any the losses of energy appears, termed the probability (fraction) of
a recoilless processes.}

For example, when $\,E_{R}<<h\nu_{ph}\;(\nu_{ph}$ - the mean frequency of
phonons), then the mean energy of recoil:%

\begin{equation}
E_{R}=(1-f)h\nu_{ph}\tag{3.4}%
\end{equation}
Hence, the probability of recoilless effect is%

\begin{equation}
f=1-{\frac{E_{R}}{h\nu_{ph}}}\tag{3.5}%
\end{equation}
According to eq.(3.3) the decrease of the recoil energy $E_{R}$ of an atom in
the structure of the lattice is related to increase of its effective mass
$(M$). In our model $M$ corresponds to the mass of the effecton.

The effect of $\gamma$-quantum irradiation without recoil was discovered by
M\"ossbauer in 1957 and named after him.

The value of M\"ossbauer effect is determined by the value of\ $f\le1$.

The big recoil energy may be transferred to the lattice by portions that are
\textbf{resonant }to the frequency of IR photons (tr and lb) and phonons.
\textbf{The possibility of superradiation of IR quanta stimulation as a result
of such recoil process is a consequence of our mesoscopic model. }

The scattering of $\gamma$-quanta without lattice excitation, when\ $E_{R}%
<<h\nu_{ph}$, is termed the \textit{elastic }one\textit{. }The general
expression [11, 12] for the probability of such phononless elastic $\gamma
$-quantum radiation acts is equal to:%

\begin{equation}
f=\exp\left(  -{\frac{4\pi<x^{2}>}{\lambda_{0}^{2}}}\right) \tag{3.6}%
\end{equation}

where $\lambda_{0}=c/\nu_{0}$ is the real wavelength of $\gamma$-quantum;
$<$x$^{2}>$ - the nucleus oscillations mean amplitude squared in the direction
of $\gamma$-quantum irradiation.

The $\gamma$\textbf{-}quanta wavelength parameter may be introduced like:%

\begin{equation}
L_{0}=\lambda_{0}/2\pi,\tag{3.7}%
\end{equation}
where: $L_{0}=1.37\cdot10^{-5}cm$ for $Fe^{57}$, then eq.(3.6) could be
written as follows:%

\begin{equation}
f=\exp\left(  -{\frac{<x^{2}>}{L_{0}^{2}}}\right) \tag{3.8}%
\end{equation}

It may be shown [12], proceeding from the model of crystal as a system of 3N
identical quantum oscillators, that when temperature (T) is much lower than
the Debye one $(\theta_{D})$ then:%

\begin{equation}
<x^{2}>={\frac{9\hbar^{2}}{4Mk\theta_{D}}}\left\{  1+{\frac{2\hbar^{2}T^{2}%
}{3\theta_{D}^{2}}}\right\}  ,\tag{3.9}%
\end{equation}
where\ $\theta_{D}=h\nu_{D}/k$ \thinspace and\thinspace$\nu_{D}\,$ is the
Debye frequency.

From (3.1), (3.3) and (3.7) we have:%

\begin{equation}
{\frac{1}{L}}={\frac{E_{0}}{\hbar c}}\tag{3.10}%
\end{equation}
where: $E_{0}=h\nu=c(2ME_{R}$)$^{1/2}\;$ is the energy of $\gamma$-quantum

Substituting eqs.(3.9 and 3.10) into eq.(3.8), we obtain the Debye-Valler formula:%

\begin{equation}
f=\exp\left[  -{\frac{E_{R}}{k\theta_{D}}}\left\{  {\frac{3}{2}}+{\frac
{\pi^{2}T^{2}}{\theta_{D}}}\right\}  \right] \tag{3.11}%
\end{equation}

when $T\rightarrow0$, then%

\begin{equation}
f\rightarrow\exp\left(  -{\frac{3E_{R}}{2k\theta_{D}}}\right) \tag{3.12}%
\end{equation}

\smallskip

\begin{center}
\smallskip
\end{center}

\smallskip

\begin{center}
{\large 3.2. Probability of elastic effects}
\end{center}

\smallskip

Mean square displacements $<$x$^{2}>$ of an atoms or molecules in condensed
matter (eq. 3.8) is not related to excitation of thermal photons or phonons
(i.e. primary or secondary transitons). According to our concept, $<x^{2}>$ is
caused by the mobility of the atoms forming effectons and differs for primary
and secondary translational and librational effectons in $(a,\bar{a})_{tr,lb}$
and $(b,b)_{tr,lb}$ states.

We will ignore below the contributions of macro- and supereffectons in
M\"{o}ssbauer effect as very small. Then the resulting probability of elastic
effects at $\gamma$-quantum radiation is determined by the sum of the
following contributions (see $eqs.4.2-4.4\,\,$of [1, 2]$)$:%

\begin{equation}
f={\frac{1}{Z}}\sum_{tr,lb}^{{}}\left[  \left(
\begin{array}
[c]{c}%
P_{ef}^{a}f_{ef}^{a}+P_{ef}^{b}f_{ef}^{b}%
\end{array}
\right)  +\left(
\begin{array}
[c]{c}%
\bar{P}_{ef}^{a}\bar{f}_{ef}^{a}+\bar{P}_{ef}^{b}\bar{f}_{ef}^{b}%
\end{array}
\right)  \right]  _{tr,lb}\tag{3.13}%
\end{equation}

where: $P_{ef}^{a},\;P_{ef}^{b},\;\bar P_{ef}^{a},\;\bar P_{ef}^{b}$ are the
relative probabilities of the \textit{acoustic and optic }states for primary
and secondary effectons; Z is the total partition function.

These parameters are calculated as described in Chapter 4 of book [1] and in
papers cited in the Summary of this article. Each of contributions to
resulting probability of the elastic effect can be calculated separately as:%

\begin{equation}
\left(
\begin{array}
[c]{c}%
f_{ef}^{a}%
\end{array}
\right)  _{tr,lb}=\exp\left[  -{\frac{<\left(
\begin{array}
[c]{c}%
x^{a}%
\end{array}
\right)  _{tr,lb}^{2}>}{L_{0}^{2}}}\right] \tag{3.14}%
\end{equation}
$\left(
\begin{array}
[c]{c}%
f_{ef}^{a}%
\end{array}
\right)  _{tr,lb}$ is the probability of elastic effect, related to dynamics
of primary translational and librational effectons in \textit{a}-state;%

\begin{equation}
\left(
\begin{array}
[c]{c}%
f_{ef}^{b}%
\end{array}
\right)  _{tr,lb}=\exp\left[  -{\frac{<\left(
\begin{array}
[c]{c}%
x^{b}%
\end{array}
\right)  _{tr,lb}^{2}>}{L_{0}^{2}}}\right] \tag{3.15}%
\end{equation}
$\left(
\begin{array}
[c]{c}%
f_{ef}^{b}%
\end{array}
\right)  _{tr,lb}$ is the probability of elastic effect in primary
translational and librational effectons in \textit{b}-state;%

\begin{equation}
\left(
\begin{array}
[c]{c}%
\bar{f}_{ef}^{a}%
\end{array}
\right)  _{tr,lb}=\exp\left[
\begin{array}
[c]{c}%
-{\frac{<\left(
\begin{array}
[c]{c}%
\bar{x}^{a}%
\end{array}
\right)  _{tr,lb}^{2}>}{L_{0}^{2}}}%
\end{array}
\right] \tag{3.16}%
\end{equation}
$\left(
\begin{array}
[c]{c}%
\bar{f}_{ef}^{a}%
\end{array}
\right)  _{tr,lb}$ is the probability for secondary effectons in \textit{\={a}
}-state;%

\begin{equation}
\left(
\begin{array}
[c]{c}%
\bar{f}_{ef}^{b}%
\end{array}
\right)  _{tr,lb}=\exp\left[  -{\frac{<\left(
\begin{array}
[c]{c}%
\bar{x}^{b}%
\end{array}
\right)  _{tr,lb}^{2}>}{L_{0}^{2}}}\right] \tag{3.17}%
\end{equation}
$\left(
\begin{array}
[c]{c}%
\bar{f}_{ef}^{b}%
\end{array}
\right)  _{tr,lb}$ is the probability of elastic effect, related to secondary
effectons in \textit{\={b}}-state.

Mean square displacements within different types of effectons in
eqs.(3.14-3.17) are related to their phase and group velocities. At first we
express the displacements using group velocities of the waves $B(v_{gr})$ and
periods of corresponding oscillations $(T\,)$ as:%

\begin{equation}
<\left(
\begin{array}
[c]{c}%
x^{a}%
\end{array}
\right)  _{tr,lb}^{2}>\,={\frac{<(v_{gr}^{a})_{tr,lb}^{2}>}{<\nu_{a}%
^{2}>_{tr,lb}}}=\,<\left(
\begin{array}
[c]{c}%
v_{gr}^{a}T^{a}%
\end{array}
\right)  _{tr,lb}^{2}>\tag{3.18}%
\end{equation}
where $(T^{a})_{tr,lb}=(1/\nu_{a})_{tr,lb}$ is a relation between the period
and the frequency of primary translational and librational effectons in
\textit{a}-state;

$(v^{a}_{gr} = v^{b}_{gr})_{tr,lb}$ are the group velocities of atoms forming
these effectons equal in (a) and (b) states.

In a similar way we can express the displacements of atoms forming (b) state
of primary effectons (tr and lib):%

\begin{equation}
<\left(
\begin{array}
[c]{c}%
x^{b}%
\end{array}
\right)  _{tr,lb}^{2}>={\frac{<(v_{gr}^{b})_{tr,lb}^{2}>}{<\nu_{b}%
^{2}>_{tr,lb}}}\tag{3.19}%
\end{equation}
where $\nu_{b}$ is the frequency of \textit{primary }translational and
librational effectons in \textit{b}-state.

The mean square displacements of atoms forming \textit{secondary
}translational and librational effectons in \textit{\={a} }and \textit{\={b} }states:%

\begin{equation}
<\left(
\begin{array}
[c]{c}%
\bar{x}^{a}%
\end{array}
\right)  _{tr,lb}^{2}>={\frac{<(\overline{v}_{gr}^{a})_{tr,lb}^{2}>}{<\bar
{\nu}_{a}^{2}>_{tr,lb}}}\tag{3.20}%
\end{equation}%
\begin{equation}
<\left(
\begin{array}
[c]{c}%
\bar{x}^{b}%
\end{array}
\right)  _{tr,lb}^{2}>={\frac{<(\overline{v}_{gr}^{b})_{tr,lb}^{2}>}{<\bar
{\nu}_{b}^{2}>_{tr,lb}}}\tag{3.21}%
\end{equation}%
\[
\text{where:\ }(\bar{v}_{gr}^{a}=\bar{v}_{gr}^{b})_{tr,lb}
\]
Group velocities of atoms in primary and secondary effectons may be expressed
using corresponding phase velocities $(v_{ph})$ and formulae for waves B
length as follows:%

\begin{align}
\left(
\begin{array}
[c]{c}%
\lambda_{a}%
\end{array}
\right)  _{tr,lb}  & ={\frac{h}{m<v_{gr}>_{tr,lb}}}=\left(  {\frac{v_{ph}^{a}%
}{\nu_{a}}}\right)  _{tr,lb}=\tag{3.22}\\
& =\left(
\begin{array}
[c]{c}%
\lambda_{b}%
\end{array}
\right)  _{tr,lb}=\left(  {\frac{v_{ph}^{b}}{\nu_{b}}}\right)  _{tr,lb}%
\nonumber
\end{align}
hence for the group velocities of the atoms or molecules forming primary
effectons (\textit{tr and lb}) squared we have:%

\begin{equation}
\left(
\begin{array}
[c]{c}%
v_{gr}^{a,b}%
\end{array}
\right)  _{tr,lb}^{2}={\frac{h^{2}}{m^{2}}}\left(  {\frac{\nu_{a,b}}%
{v_{ph}^{a,b}}}\right)  _{tr,lb}^{2}\tag{3.23}%
\end{equation}
In accordance with mesoscopic theory, the wave B length, impulses and group
velocities in \textit{a }and \textit{b }states of the effectons are equal.
Similarly to (3.23), we obtain the group velocities of particles, composing
secondary effectons:%

\begin{equation}
\left(
\begin{array}
[c]{c}%
\bar{v}_{gr}^{a,b}%
\end{array}
\right)  _{tr,lb}^{2}={\frac{h^{2}}{m^{2}}}\left(  {\frac{\bar{\nu}_{a,b}%
}{\bar{v}_{ph}^{a,b}}}\right)  _{tr,lb}^{2}\tag{3.24}%
\end{equation}
Substituting eqs.(3.23) and (3.24) into (3.18-3.21), we find the important
expressions for the average coherent displacements of particles squared as a
result of their oscillations in the volume of the effectons (\textit{tr, lib)
}in both discreet states (acoustic and optic):%

\begin{equation}
<(x^{a})_{tr,lb}^{2}>=\,(h/mv_{ph}^{a})_{tr,lb}^{2}\tag{3.25}%
\end{equation}%
\begin{equation}
<(x^{b})_{tr,lb}^{2}>=(h/mv_{ph}^{b})_{tr,lb}^{2}\tag{3.26}%
\end{equation}%
\begin{equation}
<(\overline{x}^{a})_{tr,lb}^{2}>=(h/m\overline{v}_{ph}^{a})_{tr,lb}%
^{2}\tag{3.27}%
\end{equation}%
\begin{equation}
<(\overline{x}^{b})_{tr,lb}^{2}>=(h/m\overline{v}_{ph}^{b})_{tr,lb}%
^{2}\tag{3.28}%
\end{equation}
Then, substituting these values into eqs.(3.14-3.17) we obtain a set of
different contributions to the resulting probability of effects without recoil:%

\begin{equation}
\left.
\begin{array}
[c]{r}%
\left(
\begin{array}
[c]{c}%
f_{f}^{a}%
\end{array}
\right)  _{tr.lb}=\exp\left[  -\left(  {\frac{h}{mL_{0}v_{ph}^{a}}}\right)
^{2}\right]  _{tr,lb};\\
\left(
\begin{array}
[c]{c}%
f_{f}^{b}%
\end{array}
\right)  _{tr.lb}=\exp\left[  -\left(  {\frac{h}{mL_{0}v_{ph}^{b}}}\right)
^{2}\right]  _{tr,lb};
\end{array}
\right\} \tag{3.29}%
\end{equation}%
\begin{equation}
\left.
\begin{array}
[c]{r}%
\left(
\begin{array}
[c]{c}%
\bar{f}_{f}^{a}%
\end{array}
\right)  _{tr.lb}=\exp\left[  -\left(  {\frac{h}{mL_{0}\bar{v}_{ph}^{a}}%
}\right)  ^{2}\right]  _{tr,lb};\\
\left(
\begin{array}
[c]{c}%
\bar{f}_{f}^{b}%
\end{array}
\right)  _{tr.lb}=\exp\left[  -\left(  {\frac{h}{mL_{0}\bar{v}_{ph}^{b}}%
}\right)  ^{2}\right]  _{tr,lb};
\end{array}
\right\} \tag{3.30}%
\end{equation}
where the phase velocities $(v_{ph}^{a},\;v_{ph}^{b},\;\bar{v}_{ph}^{a}%
,\;\bar{v}_{ph}^{b})_{tr,lb}$ are calculated from the resulting sound velocity
and the positions of translational and librational bands in the oscillatory
spectra of matter at given temperature using eqs.2.69-2.75.\ The wavelength parameter:%

\[
L_{0}={\frac c{2\pi\nu_{0}}}={\frac{hc}{2\pi E_{0}}}=1.375\cdot10^{-11}\,m
\]
for gamma-quanta, radiated by nuclear of $Fe^{57}$, with energy:%

\[
E_{0}=14.4125\text{ kev }=2.30167\cdot10^{-8}\,\text{erg }
\]
\medskip Substituting eqs.(3.29) and (3.30) into (3.13), we find the total
probability of recoilless effects $(f_{\text{tot}})$ in the given substance.
Corresponding computer calculations for ice and water are presented on
Figs.3.1 and 3.2.

As far the second order phase transitions in general case are accompanied by
the alterations of the sound velocity and the positions of translational and
librational bands, they should also be accompanied by alterations of
f$_{\text{tot}}$ and its components.

\begin{center}%
\begin{center}
\includegraphics[
height=4.0542in,
width=4.4763in
]%
{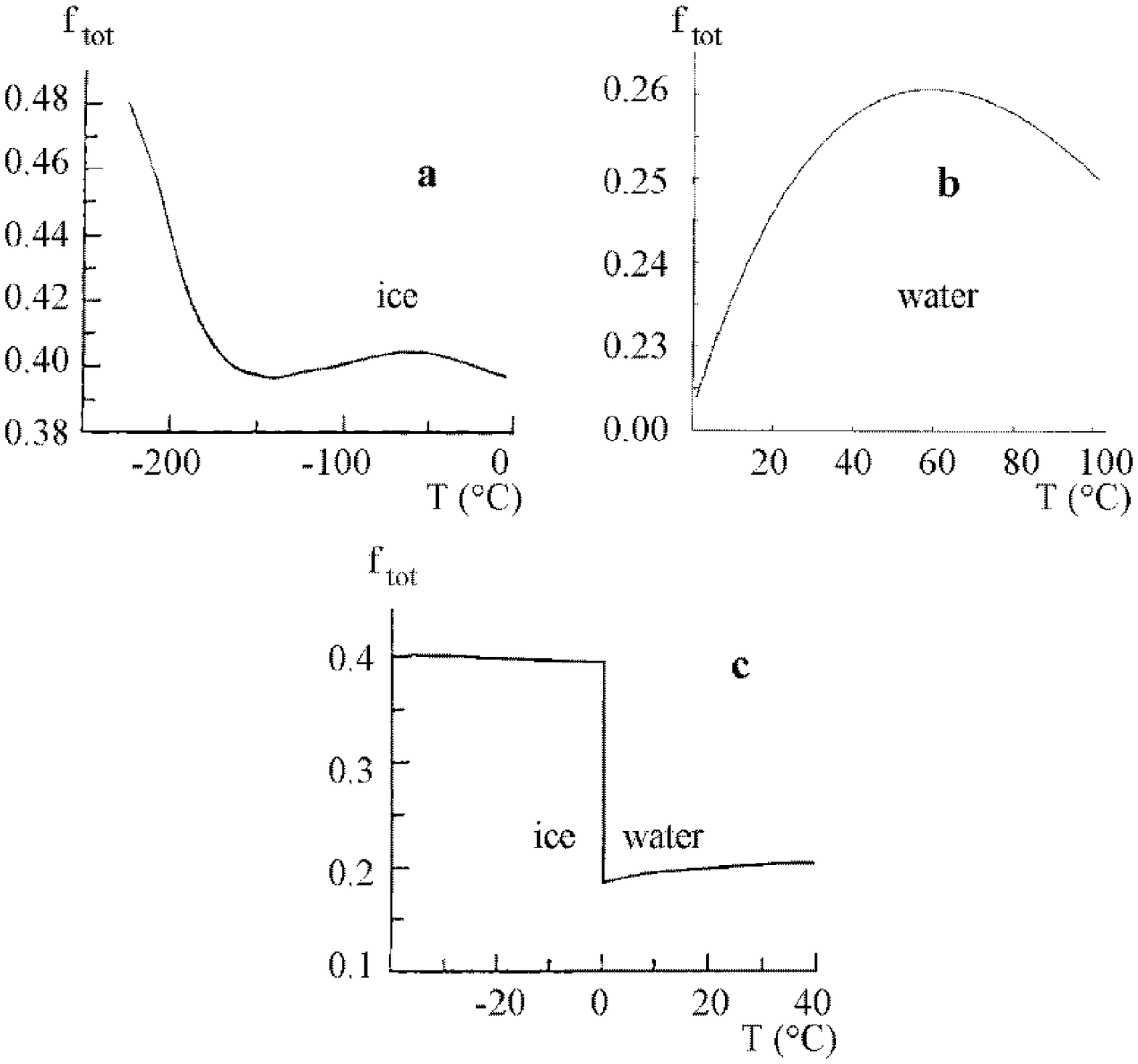}%
\end{center}
\end{center}

\begin{quotation}
\textbf{Fig.~3.1. }Temperature dependences of total probability (\textit{f})
for elastic effect without recoil and phonon excitation: (a) in ice; (b) in
water; (c)-during phase transition. The calculations were performed using
eq.(3.13).\medskip%
\begin{center}
\includegraphics[
height=1.9752in,
width=4.817in
]%
{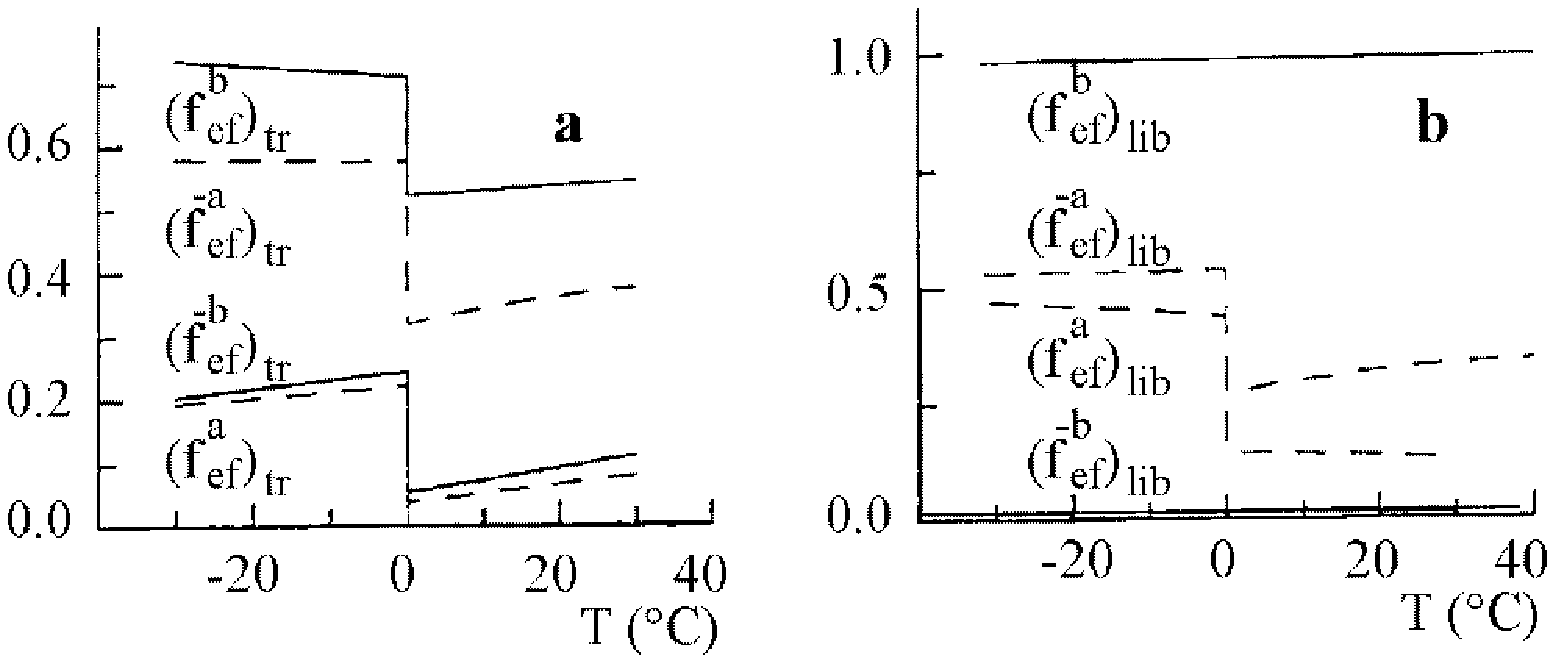}%
\end{center}

\textbf{Fig.~3.2. }(\textbf{a}) - The contributions to probability of elastic
effect (f) (see Fig.3.1) for primary $(f_{ef}^{a,b})_{tr}$ and secondary
$(\bar{f})_{tr}$ translational effectons and \textbf{(b)} and those of
librational effectons $(f_{ef}^{a,b})_{lb}$ and $(\bar{f})_{lb}$ near the
temperature of [ice $\Leftrightarrow$ water] phase transition.
\end{quotation}

\medskip

The total probability (\textit{f}) and its components, caused by primary and
secondary quasiparticles were calculated according to formula (3.13). The
value of $\left(  f\right)  $ determines the magnitude of the M\"{o}ssbauer
effect registered by $\gamma$-resonance spectroscopy.

The band width caused by recoilless effects is determined by the uncertainty
principle and expressed as follows:%

\begin{equation}
\Gamma={\frac{h}{\tau}}\approx{\frac{10^{-27}}{1.4\cdot10^{-7}}}%
=7.14\cdot10^{-21}\text{ erg }=4.4\cdot10^{-9}eV\tag{3.31}%
\end{equation}
where $\tau$ is the lifetime of nucleus in excited state (for $Fe^{57}%
\;\,\,\tau=1.4\cdot10^{-7}s)$.

The position of the band depends on the mean square velocity of atoms, i.e. on
second order Doppler effect. In the experiment, such an effect is compensated
by the \textit{velocity of }$\gamma$-\textit{quanta source }motion relative to
absorbent. In the framework of our model this velocity is interrelated with
the mean velocity of the secondary effectons diffusion in condensed matter.

\smallskip

\begin{center}
{\large 3.3. Doppler broadening in spectra of nuclear gamma-resonance (NGR)}
\end{center}

\smallskip

M\"{o}ssbauer effect is characterized by the nonbroadened component of NGR
spectra only, with probability of observation determined by eq.(3.13).

When the energy of absorbed $\gamma$-quanta exceeds the energy of thermal IR
\textit{photons (tr,lib) }or \textit{phonons excitation, }the absorbance band
broadens as a result of Doppler effect. Within the framework of our mesoscopic
concept the Doppler broadening is caused by thermal displacements of the
particles during $[a\Leftrightarrow b$ \ and $\;\bar a\Leftrightarrow\bar
b]_{tr,lb}$ transitions of primary and secondary effectons, leading to
origination/annihilation of the corresponding type of deformons
(electromagnetic and acoustic).

The \textit{flickering clusters}: $[lb/tr]$ convertons (\textit{a} and
\textit{b}), can contribute in the NGR line broadening also.

In that case, the value of Doppler broadening ($\Delta\Gamma$) of the band in
the NGR spectrum could be estimated from corresponding kinetic energies of
these excitations, related to their group velocities (see eq. 4.31). In our
consideration we take into account the \textbf{reduced to one molecule
}kinetic energies of primary and secondary translational and librational
transitons, \textit{a}-convertons and \textit{b}-convertons. The contributions
of macroconvertons, macro- and superdeformons are much smaller due to their
small probability and concentration:%

\begin{align}
\Delta\Gamma & ={\frac{V_{0}}{N_{0}Z}}\underset{tr,lb}{\sum}\left(
\begin{array}
[c]{c}%
n_{t}P_{t}T_{t}+\bar{n}_{t}\bar{P}_{t}\bar{T}_{t}%
\end{array}
\right)  _{tr,lb}+\tag{3.32}\\
& +{\frac{V_{0}}{N_{0}Z}}(n_{ef})_{lb}[P_{ac}T_{ac}+P_{bc}T_{bc}]\nonumber
\end{align}

where: $N_{0}$ and $V_{0}$ are the Avogadro number and molar volume;

Z is the total partition function $(eq.4.2);\;n_{t}$ and $\bar{n}_{t}$ are the
concentrations of primary and secondary transitons (eqs.3.5 and 3.7);

$(n_{ef})_{lb}=n_{\text{con\ }}$ is a concentration of primary librational
effectons, equal to that of the convertons; $\;P_{t}$ and\thinspace$\bar
{P}_{t}\,$ are the relative probabilities of primary and secondary transitons
(eqs. 4.26 and $4.27);\;P_{ac}$ and $P_{bc}$ are relative probabilities of
(\textit{a} and \textit{b}) -convertons (see Chapter 4 of [1]);

$T_{t}$ and $\bar{T}_{t}$ are the kinetic energies of primary and secondary
transitons, related to the corresponding total energies of these excitations
$(E_{t}$ and $\bar{E}_{t})$, their masses $(M_{t}$ and $\overline{M}_{t})$ and
the resulting sound velocity $(v_{s}$, see eq.2.40) in the following form:%

\begin{equation}
(T_{t})_{tr,lb}={\frac{\sum_{1}^{3}\left(  E_{t}^{1,2,3}\right)  _{tr,lb}%
}{2M_{t}(v_{s}^{\text{res}})^{2}}}\tag{3.33}%
\end{equation}%
\begin{equation}
(T_{t})_{tr,lb}={\frac{\sum_{1}^{3}\left(  \bar{E}_{t}^{1,2,3}\right)
_{tr,lb}}{2\bar{M}_{t}(v_{s}^{\text{res}})^{2}}}\tag{3.34}%
\end{equation}
The kinetic energies of (a and b) convertons are expressed in a similar way:%

\[
(T_{ac})={\frac{\sum_{1}^{3}\left(  E_{ac}^{1,2,3}\right)  _{tr,lb}}%
{2M_{c}(v_{s}^{\text{res}})^{2}}}
\]
\[
(T_{bc})={\frac{\sum_{1}^{3}\left(  E_{bc}^{1,2,3}\right)  _{tr,lb}}%
{2M_{c}(v_{s}^{\text{res}})^{2}}}
\]
where: $E_{ac}^{1,2,3}$ and $\;E_{bc}^{1,2,3}$ are the energies of selected
states of corresponding convertons; $M_{c}$ is the mass of convertons, equal
to that of primary librational effectons.

The broadening of NGR spectral lines by Doppler effect in liquids is generally
expressed using the diffusion coefficient (D) at the assumption that the
motion of M\"{o}ssbauer atom has the character of unlimited diffusion [13]:%

\begin{equation}
\Delta\Gamma={\frac{2E_{0}^{2}}{\hbar c^{2}}}D\tag{3.35}%
\end{equation}
where: $E_{0}=h\nu_{0}$ is the energy of gamma quanta; c is light velocity and%

\begin{equation}
D={\frac{kT}{6\pi\eta a}}\tag{3.36}%
\end{equation}
where: $\eta$ is viscosity, (a) is the effective Stokes radius of the atom
$Fe^{57}$

The probability of recoilless $\gamma$-quantum absorption by the matter
containing for example Fe$^{57}$, decreases due to diffusion and corresponding
Doppler broadening of band $(\Delta\Gamma)$:%

\begin{equation}
f_{D}={\frac{\Gamma}{\Gamma+\Delta\Gamma}}\tag{3.37}%
\end{equation}
where $\Delta\Gamma$ corresponds to eq.(3.32).The formulae obtained here make
it possible to experimentally verify a set of consequences of our mesoscopic
theory using the gamma- resonance method. A more detailed interpretation of
the data obtained by this method also becomes possible.

The magnitude of ($\Delta\Gamma$) was calculated according to formula (3.32).
It corresponds well to experimentally determined Doppler widening in the
nuclear gamma resonance (NGR) spectra of ice.

\begin{center}%
\begin{center}
\includegraphics[
height=4.0534in,
width=4.6034in
]%
{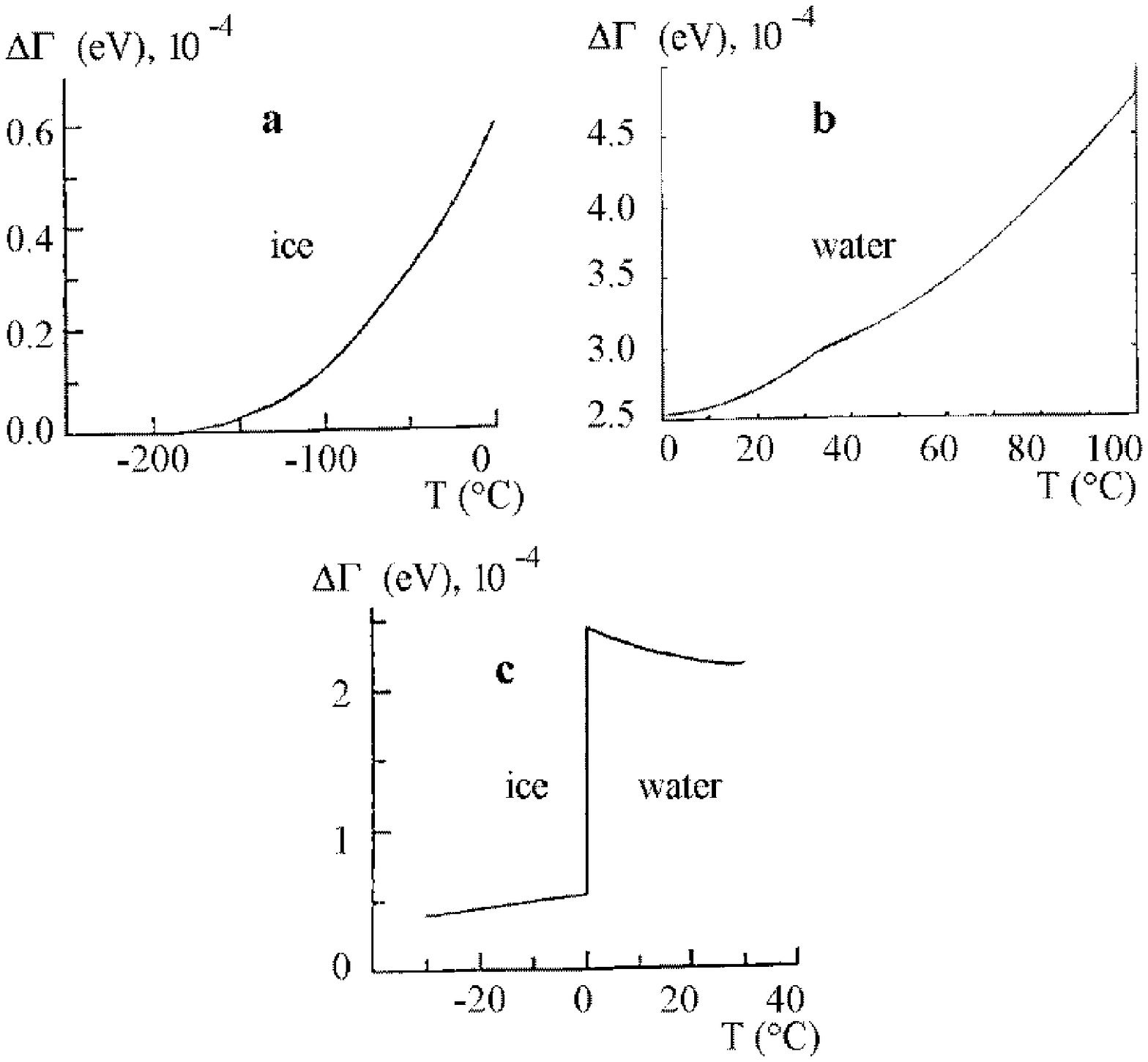}%
\end{center}
\end{center}

\begin{quotation}
\textbf{Fig.~3.3. }The temperature dependences of the parameter $\Delta\Gamma
$, characterizing the nonelastic effects and related to the excitation of
thermal phonons and IR photons: \textbf{a}) in ice; \textbf{b}) in water;
\textbf{c}) near phase transition.

\smallskip
\end{quotation}

\begin{center}
{\large 3.4. Acceleration and forces, related to thermal dynamics of molecules
and ions.}

{\large Hypothesis of Vibro-gravitational interaction}
\end{center}

{\large \smallskip}

During the period of particles thermal oscillations (tr and lb), their instant
velocity, acceleration and corresponding forces alternatively and strongly change.

The change of wave B instant group velocity, averaged during the molecule
oscillation period in composition of the (a) and (b) states of the effectons,
determines the average \textit{acceleration: }%

\begin{equation}
\left[  a_{gr}^{a,b}={\frac{dv_{gr}^{a,b}}{dt}}={\frac{v_{gr}^{a,b}}{T}%
}=v_{gr}\nu^{a,b}\right]  _{tr,lb}^{1,2,3}\tag{3.38}%
\end{equation}
We keep in mind that group velocities, impulses and wave B length in (a) and
(b) states of the effectons are equal, in accordance with our model.

Corresponding to (3.38) forces:%

\begin{equation}
\left[  F^{a,b}=ma_{gr}^{a,b}\right]  _{tr,lb}^{1,2,3}\tag{3.39}%
\end{equation}
The energies of molecules in (a) and (b) states of the effectons also can be
expressed via accelerations:%

\begin{equation}
\left[  E^{a,b}=h\nu^{a,b}=F^{a,b}\lambda=ma^{a,b}\cdot\lambda=ma^{a,b}%
(v_{ph}^{a,b}/\nu^{a,b})\right]  _{tr,lb}^{1,2,3}\tag{3.40}%
\end{equation}
From (3.40) one can express the accelerations of particles in the primary
effectons of condensed matter, using their phase velocities as a waves B:%

\begin{equation}
\left[  a_{gr}^{a,b}={\frac{h(\nu^{a,b})^{2}}{mv_{ph}^{a,b}}}\right]
_{tr,lb}^{1,2,3}\tag{3.41}%
\end{equation}
The accelerations of particles in composition of secondary effectons have a
similar form:%

\begin{equation}
\left[  \bar{a}_{gr}^{a,b}={\frac{h(\bar{\nu}^{a,b})^{2}}{m\bar{v}_{ph}^{a,b}%
}}\right]  _{tr,lb}^{1,2,3}\tag{3.42}%
\end{equation}
These parameters are important for understanding the dynamic properties of
condensed systems. The accelerations of the atoms, forming primary and
secondary effectons can be calculated, using eqs.(2.74-2.75 of [1]) to
determine phase velocities and eqs. $(2.27,\;2.28,\;2.54,\;2.55\,\,\,[1])$ to
find a frequencies.

Multiplying (3.41) and (3.42) by the atomic mass $\,\,m$, we derive the most
probable and mean forces acting upon the particles in both states of primary
and secondary effectons in condensed matter:%

\begin{equation}
\left[  F_{gr}^{a,b}={\frac{h(\nu^{a,b})^{2}}{v_{ph}^{a,b}}}\right]
_{tr,lb}^{1,2,3}\;\;\;\;\;\;\;\;\left[  \bar{F}_{gr}^{a,b}={\frac{h(\bar{\nu
}^{a,b})^{2}}{\bar{v}_{ph}^{a,b}}}\right] \tag{3.43}%
\end{equation}
The comparison of calculated accelerations with empirical data of the
M\"{o}ssbauer effect - supports the correctness of our approach.

According to eq.(2.54) in the low temperature range, when $h\nu_{a}<<kT$, the
frequency of \textit{secondary tr and lb} effectons in the (a) state can be
estimated as:%

\begin{equation}
\nu^{a}={\frac{\nu_{a}}{\exp\left(  {\frac{h\nu_{a}}{kT}}\right)  -1}}%
\approx{\frac{kT}{h}}\tag{3.44}%
\end{equation}
For example, at $T=200K$ \ we have\ $\bar{\nu}^{a}\approx4\cdot10^{12}s^{-1}$.

If the phase speed in eq.(3.42) is taken equal to $\bar{v}_{ph}^{a}%
=2.1\cdot10^{5}cm/s\;($see Fig.2a) and the mass of water molecule:%

\[
m=18\cdot1.66\cdot10^{-24}g=2.98\cdot10^{-23}g,
\]
then from (3.42) we get the acceleration of molecules in composition of
secondary effectons of ice in (a) state:%

\[
\bar{a}_{gr}^{a}={\frac{h(\bar{\nu}^{a})^{2}}{m\bar{v}_{ph}^{a}}}%
=1.6\cdot10^{16}cm/s^{2}
\]
This value is about $10^{13}$ times more than that of free fall acceleration
$(g=9.8\cdot10^{2}cm/s^{2})$, which agrees well with experimental data,
obtained for solid bodies [11].

Accelerations of $H_{2}O$ molecules in composition of \textit{primary
librational }effectons $(a_{gr}^{a})$ in the ice at 200K and in water at 300K
are equal to: $0.6\cdot10^{13}cm/s^{2}\;$ and\ $2\cdot10^{15}cm/s^{2}$,
correspondingly. \textbf{They also exceed to many orders the free fall acceleration.}

It was shown experimentally (Sherwin, 1960), that heating of solid body leads
to decreasing of gamma-quanta frequency (red Doppler shift) i.e. increasing of
corresponding quantum transitions period. This can be explained as the
relativist time-pace decreasing due to elevation of average thermal velocity
of atoms [11].

The thermal vibrations of particles (atoms, molecules) in composition of
primary effectons as a partial Bose-condensate are coherent. The increasing of
such clusters dimensions, determined by most probable wave B length, as a
result of cooling, pressure elevation or under magnetic field action (see
section 14.6 of [1]) leads to enhancement of coherent regions.

\textbf{Each coherently vibrating cluster of particles with big alternating
accelerations, like librational and translational effectons is a source of
coherent gravitational waves. }

The frequency of these vibro-gravitational waves (VGW) is equal to frequency
of particles vibrations (i.e. frequency of the effectons in \textit{a }or
\textit{b }states). The amplitude of VGW\ $(A_{G})$ is proportional to the
number of vibrating coherently particles $(N_{G})$ in composition of primary effectons:%

\[
A_{G}\sim N_{G}\sim V_{ef}/(V_{0}/N_{0})=(1/n_{ef})/(V_{0}/N_{0})
\]
\textbf{The resonant long-distance gravitational interaction between coherent
clusters of the same body or between that of different bodies is possible}.
\textbf{The formal description of this vibro-gravitational interaction (VGI)
could be like that of distant macroscopic Van der Waals interaction.}

\textbf{Different patterns of nonlocal Bose-condensate of standing
gravitational waves in vacuum represent the field-informational copy of local
Bose- condensate of the effectons of condensed matter.}

\textbf{Very important role of proposed here distant resonant
VIBRO-GRAVITATIONAL INTERACTION (VGI) in elementary acts of perception and
memory can be contributed by coherent primary librational water effectons in
microtubules of the nerve cells (see paper ''Hierarchic Model of
Consciousness'' in URL: http://www.karelia.ru/\symbol{126}alexk \ and \ \ http://arXiv.org/abs/physics/0003045).}

\bigskip

\begin{quotation}
\bigskip

{\Large \ \ \ \ References }

\smallskip

\smallskip

\textbf{[1]. Kaivarainen A. Hierarchic Concept of Matter and Field. Water,
biosystems and elementary particles. New York, 1995, pp. 485.}

\textbf{[2]. \thinspace Kaivarainen A. New Hierarchic Theory of Matter General
for Liquids and Solids: dynamics, }\thinspace\textbf{thermodynamics and
mesoscopic structure of water and ice }

\textbf{(see URL: http://www.karelia.ru/\symbol{126}alexk) }

[\textbf{3]. Kaivarainen A. \thinspace Hierarchic Concept of Matter, General
for Liquids and Solids: Water and ice \thinspace}\thinspace\textbf{(see
Proceedings of the Second Annual Advanced Water Sciences Symposium,
}\thinspace\textbf{October 4-6, 1966, Dallas, Texas.}

\textbf{[4]. Eisenberg D., Kauzmann W. The structure and properties of water.
Oxford University Press, Oxford, 1969.}

\textbf{[5]. Frontas'ev V.P., Schreiber L.S. J. Struct. Chem. (USSR}%
$)\,\,$\textbf{6,  (1966), 512}$.$

\textbf{[6]. Kikoin I.K. (Ed.) Tables of physical values. Atomizdat, Moscow,
1976 (in Russian).}

\textbf{[7]. Einstein A. Collection of works. Nauka, Moscow, 1965.}

\textbf{[8]. Vuks M.F. Light scattering in gases, liquids and solutions.
Leningrad University Press, Leningrad. 1977.}

\textbf{[9]. Vuks M.F. Electrical and optical properties of molecules and
condensed matter. Leningrad University Press, Leningrad, 1984.}

\textbf{[10]. Theiner O., White K.O. J.Opt. Soc.Amer. }$\;1969,\,59,\,181$\textbf{.}

\textbf{[11]. Wertheim G.K. M\"{o}ssbauer effect. Academic Press, N.Y. and
London. 1964.}

\textbf{[12]. Shpinel V.C. Gamma-ray resonance in crystals. Nauka, Moscow, 1969.}

\textbf{[13]. Singvi K., Sielander A. In book: M\"{o}ssbauer effect. Ed. Kogan
Yu. Moscow 1962.}
\end{quotation}
\end{document}